\documentstyle[epsfig]{elsart}
\begin{document}

\newcommand{\re}{\mathop{\mathrm{Re}}}
\newcommand{\im}{\mathop{\mathrm{Im}}}
\newcommand{\I}{\mathop{\mathrm{i}}}
\newcommand{\D}{\mathop{\mathrm{d}}}
\newcommand{\E}{\mathop{\mathrm{e}}}

\def\lambar{\lambda \hspace*{-5pt}{\rule [5pt]{4pt}{0.3pt}} \hspace*{1pt}}

{\Large  DESY 12-070}

{\Large  May 2012}

\bigskip

\begin{frontmatter}

\journal{Phys. Rev. ST-AB}

\date{}

\title{
Harmonic lasing in X-ray FELs}

\author{E.A.~Schneidmiller}
\author{and M.V.~Yurkov}

\address{Deutsches Elektronen-Synchrotron (DESY),
Notkestrasse 85, D-22607 Hamburg, Germany}

\begin{abstract}
Harmonic lasing in a free electron laser with a planar undulator
(under the condition that the fundamental frequency is suppressed) might be a cheap and efficient way of extension of wavelength ranges of
existing and planned X-ray FEL facilities. Contrary to nonlinear harmonic generation,
harmonic lasing can provide much more intense, stable, and narrow-band FEL beam
which is easier to handle due to the suppressed
fundamental frequency. In this paper we perform a parametrization of the solution of the
eigenvalue equation for lasing at odd harmonics, and present an explicit expression for FEL gain length,
taking into account all essential effects. We propose and discuss methods for suppression of the fundamental harmonic.
We also suggest a combined use of harmonic lasing and lasing at the retuned fundamental wavelength
in order to reduce bandwidth and to
increase brilliance of X-ray beam at saturation.
Considering 3rd harmonic lasing as a practical example, we come to the conclusion that it is much more robust
than usually thought, and can be widely used in the existing or planned X-ray FEL facilities.
In particular, LCLS after a minor modification can lase to saturation at the 3rd harmonic up to the photon energy
of 25-30 keV providing multi-gigawatt power level and narrow bandwidth.
As for the European XFEL, harmonic lasing would allow to extend operating range (ultimately up to 100 keV),
to reduce FEL bandwidth and to increase brilliance,
to enable two-color operation for pump-probe experiments,
and to provide more flexible operation at different electron energies.
Similar improvements can be realized in other X-ray FEL facilities with gap-tunable undulators
like FLASH II, SACLA, LCLS II, etc.
Harmonic lasing can be an attractive option for compact X-ray FELs
(driven by electron beams with a relatively low energy),
allowing the use of the standard undulator technology instead of small-gap in-vacuum devices.
Finally, in this paper we discover that in a part of the parameter
space, corresponding to the operating range of soft X-ray beamlines of X-ray FEL facilities (like SASE3 beamline of the
European XFEL), harmonics can grow faster than the fundamental wavelength. This feature can be used in
some experiments, but might also be an unwanted phenomenon, and we discuss possible measures to diminish it.
\end{abstract}

\end{frontmatter}

\baselineskip 20pt

\clearpage

\section{Introduction}

Successful operation of X-ray free electron lasers (FELs) \cite{flash,lcls,sacla}, based on
self-amplified spontaneous emission (SASE) principle \cite{ks-sase},
down to an $\rm{\AA}$ngstr{\"o}m regime opens up new horizons for photon science. Even shorter
wavelengths are requested by the scientific community.
A possible way to extend operating range of a high-gain FEL is to use nonlinear harmonic generation
\cite{flash,hg-2,hg-3,kim-1,harm-prst,hg-exp-1,lcls-harm}
when bunching at harmonics is driven by the fundamental frequency in the vicinity of saturation. Then odd harmonics can be
radiated in the same (planar) undulator. However, intensity of
harmonics is rather small, for example the third one is typically at the level of a per cent of the fundamental harmonic
intensity \cite{flash,hg-3,kim-1,harm-prst,lcls-harm},
and higher harmonics are much weaker.
In addition,
for a typical user experiment one has to suppress the fundamental frequency by external filters what might also result in an additional
suppression of the harmonic intensity. Note also that a relative bandwidth of a harmonic is approximately the same as that of the
fundamental mode \cite{harm-prst} contrary to the incoherent undulator radiation for which it is inversely proportional to a harmonic number.
Finally, nonlinear harmonic generation in a SASE FEL is more strongly subjected to fluctuations than lasing at the
fundamental wavelength \cite{kim-1,harm-prst,atto-stat}.

An alternative option is a harmonic lasing that was first proposed for FEL oscillators \cite{colson},
and was experimentally demonstrated in infrared and optical wavelength ranges for oscillator
configurations \cite{benson-madey,warren,hajima,sei}.
Harmonic lasing in single-pass high-gain FELs \cite{hg-2,kim-1,murphy,mcneil},
i.e. the radiative instability at an odd harmonic of the
planar undulator developing independently from lasing at the fundamental wavelength, might have significant advantages over
nonlinear harmonic generation (much higher power, much better stability, smaller bandwidth and no necessity in filters),
provided that lasing at the fundamental frequency is suppressed
(for lasing at the 3rd harmonic)\footnote{If one is going to lase at the 5th harmonic, then both the fundamental wavelength
and the 3rd harmonic must be suppressed, and so on.}.

A possible method to suppress the fundamental harmonic without affecting the
third harmonic lasing was suggested in \cite{mcneil}: one can
use $2\pi/3$ phase shifters between undulator modules. We found out, however, that this method is
inefficient in the case of a SASE FEL (the simulations in \cite{mcneil} were done for the case of a monochromatic seed).
In this paper we suggest a modification of the phase shifters method which can also work in the case of a SASE FEL.
We also propose suppression of the fundamental harmonic by using a spectral filter in a chicane installed between two parts of
the undulator (one can also use a closed bump formed by movable quadrupoles of the undulator focusing system).
Such a chicane, for example, is used in Linac Coherent Light Source (LCLS) \cite{lcls}
as a part of the self-seeding scheme \cite{geloni}.
In either case only minor or no modifications of existing or planned
undulator systems are required so that harmonic lasing can be considered as a (practically) free option.

The key question, however, is whether or not the harmonic lasing is sufficiently robust with respect to the electron beam and undulator
quality. Undulators for X-ray FELs are usually designed and
built with a sufficient safety margin in terms of length and quality \cite{nuhn}. As for
the electron beam quality, there is a general
opinion that harmonic lasing is too sensitive to emittance and energy spread effects,
so that it is not practically interesting.
One of the main goals of this paper is to disprove this statement.
We study a realistic three-dimensional (3D) model of harmonic lasing and compare its gain length with
that of the fundamental mode. We find out that harmonic lasing
is of interest in many practical cases.

In order to calculate FEL gain length (and, therefore, saturation length) one has to solve an eigenvalue equation.
Eigenvalue equation for harmonic lasing was derived in the framework of one-dimensional (1D) model in \cite{hg-2,murphy},
and a thorough 1D analysis can be found in \cite{mcneil}.
Usually, more realistic 3D model is required to
make conclusions on a possibility of practical realization of some option. Three-dimensional analysis was done in \cite{kim-1},
where an eigenvalue equation was derived based on an approach developed in \cite{ming} for the fundamental frequency.
However, this eigenvalue equation
is rather complicated and can be solved only numerically. One can correctly calculate the gain length for a specific
set of parameters,
but it is very difficult to trace general dependencies and perform analysis of the parameter space.

In this paper  we perform a parametrization of the solution of the
eigenvalue equation for lasing at odd harmonics \cite{kim-1}, and present explicit (although approximate)
expressions for FEL gain length,
optimal beta-function, and saturation length taking into account emittance, betatron motion, diffraction of radiation,
energy spread and its growth along the undulator length due to quantum fluctuations of the undulator radiation.
Considering 3rd harmonic lasing as
a practical example, we come to the conclusion that it is much more robust
than usually thought, and can be widely used at the
present level of accelerator and FEL technology. We surprisingly find out that in many cases the 3D model of harmonic
lasing gives more optimistic results than the 1D model. For instance, one of the results of our studies is that
in a part of the parameter
space, corresponding to the operating range of soft X-ray beamlines of X-ray FEL facilities,
harmonics can grow faster than the fundamental mode.

We briefly discuss properties of saturated harmonic lasing,
and conclude that at a given wavelength the brilliance of a harmonic is approximately the same as that of the
retuned fundamental mode.
We suggest a combined use of harmonic lasing and lasing at the same wavelength with
the retuned fundamental mode in order to reduce bandwidth and to
increase brilliance of X-ray beam at saturation.

We consider a possible application of harmonic lasing to different X-ray FEL facilities, and conclude
that they can strongly profit from this option. In particular, LCLS \cite{lcls}
can significantly extend its operating range towards shorter wavelengths making use of the third harmonic lasing
with the help of the intra-undulator spectral filtering and phase shifters.
In the case of the European XFEL \cite{euro-xfel-tdr}, the harmonic lasing can
allow to extend the operating range, to reduce FEL bandwidth and increase brilliance,
to enable two-color operation for pump-probe experiments,
and to provide more flexible operation at different electron energies. Similar
improvements can be realized in other X-ray FEL facilities with gap-tunable undulators
like FLASH II \cite{flash2}, SACLA \cite{sacla}, LCLS II \cite{lcls2}, etc.
Finally, let us mention that the results of this paper can also be
used for high-gain FELs using external seed (if, for example, the 3rd or the 5th harmonic of the undulator is tuned to
the seed frequency).

\section{Gain length of harmonic lasing}

The results of this Section are generalizations of the results of Ref.~\cite{des-form} for the fundamental frequency
to the case of harmonic lasing.
The eigenvalue equation \cite{kim-1} and the approach to its parametrization are discussed in Appendix A.

Let us consider an axisymmetric electron beam with a current $I$, and a Gaussian distribution
in transverse phase space and in energy. The resonance condition for the fundamental wavelength is written as:

\begin{equation}
\lambda_1 = \frac{\lambda_{\mathrm{w}}(1+K^2)}{2 \gamma^2} \ .
\label{resonance}
\end{equation}

More generally, lasing in a planar undulator can be achieved at the odd harmonics  defined by the condition

\begin{displaymath}
\lambda_h = \frac{\lambda_1}{h} \ , \ \ \ h = 1,3,5, ...
\end{displaymath}

\noindent Here $\lambda_{\mathrm{w}}$ is the undulator period, $\gamma$ is relativistic
factor, and $K$ is the {\bf rms} undulator parameter:

\begin{equation}
K = 0.934 \ \lambda_{\mathrm{w}} [{\mathrm{cm}}] \ B_{\mathrm{rms}} [{\mathrm{T}}] \ ,
\label{k-rms}
\end{equation}

\noindent $B_{\mathrm{rms}}$ being the rms undulator field.

In what follows we assume that
the harmonic with a number $h$ lases to saturation, while lasing at harmonics with lower numbers and at the
fundamental wavelength
is suppressed with the help of phase shifters or by other means (see Section 4). We also assume that
the beta-function is optimized so that the FEL gain length
at a considered harmonic
achieves the minimum for given wavelength, beam and undulator parameters.
Under this condition the solution of the eigenvalue equation for the {\bf field} gain
length\footnote{e-folding length for the field amplitude.
There is also a notion of the power gain length which is twice shorter, see Appendix C.} of the $TEM_{00}$ mode
can be approximated as follows (see Appendix A for details):

\begin{equation}
L_g \simeq L_{g0} \ (1+\delta) \ ,
\label{lg}
\end{equation}

\noindent where

\begin{equation}
L_{g0} = 1.67 \left(\frac{I_A}{I} \right)^{1/2} \frac{(\epsilon_n \lambda_{\mathrm{w}})^{5/6}}
{\lambda_h^{2/3}} \ \frac{(1+K^2)^{1/3}}{h^{5/6} K A_{JJh}} \ ,
\label{lg0}
\end{equation}

\noindent and

\begin{equation}
\delta = 131 \ \frac{I_A}{I} \ \frac{\epsilon_n^{5/4}}
{\lambda_h^{1/8} \lambda_{\mathrm{w}}^{9/8}} \ \frac{h^{9/8} \sigma_{\gamma}^2}{(K A_{JJh})^2 (1+K^2)^{1/8}} \ .
\label{delta}
\end{equation}

\begin{figure*}[tb]

\includegraphics[width=.8\textwidth]{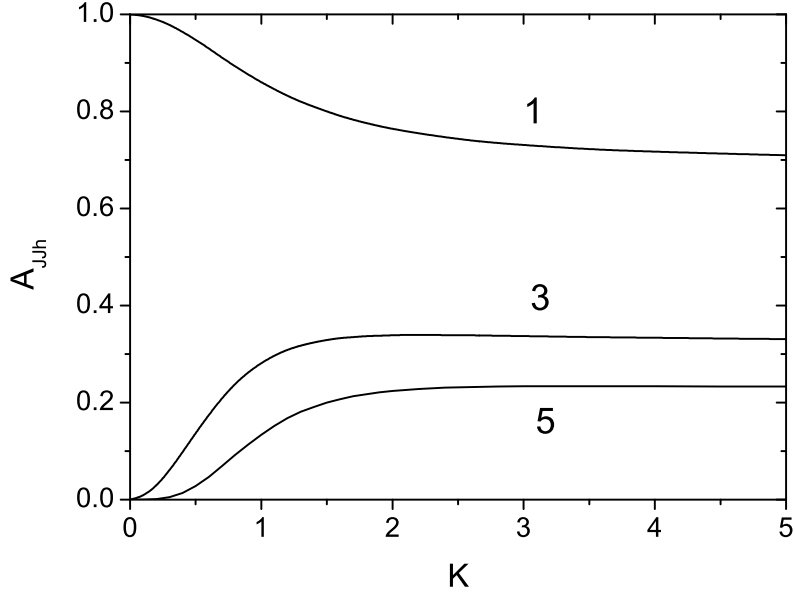}

\caption{\small Coupling factors for the 1st, 3rd, and 5th harmonics (denoted with 1, 3, and 5, correspondingly) versus
rms undulator parameter.}

\label{ajj}
\end{figure*}

\noindent The following notations are introduced here: $I_A = 17$ kA is the Alfven current,
$\epsilon_n = \gamma \epsilon$ is the rms normalized emittance,
$\sigma_{\gamma}=\sigma_{_{\cal E}}/mc^2$ is the rms energy spread
(in units of the rest energy), and

\begin{displaymath}
A_{JJh}(K) = J_{(h-1)/2} \left( \frac{hK^2}{2(1+K^2)} \right) - J_{(h+1)/2} \left( \frac{hK^2}{2(1+K^2)} \right)
\end{displaymath}

is the usual coupling factor for harmonics with $J_n$ being Bessel functions. The coupling factors for the 1st, 3rd, and 5th
harmonics are shown in Fig.~\ref{ajj}. When the rms undulator parameter $K$ is
large, the coupling factors are $A_{JJ1} \simeq 0.696$, $A_{JJ3} \simeq 0.326$, $A_{JJ5} \simeq 0.230$.
Asymptotically for large $h$ we have $A_{JJh} \simeq 0.652 \ h^{-2/3}$.
Also note that all the formulas of this Section
are valid in the case of helical undulator
and the fundamental wavelength ($h=1$), in this case the coupling factor is equal to 1 \cite{des-form}.

The formulas (\ref{lg})-(\ref{delta}) provide an accuracy better than 5 \% in the range of
parameters

\begin{equation}
1 < \frac{2 \pi \epsilon}{\lambda_h} < 5 \ ,
\label{emit-lam-lim}
\end{equation}
\begin{equation}
\delta < 2.5 \ \left\{ 1-
\exp \left[ - \frac{1}{2} \, \left( \frac{2 \pi \epsilon}{\lambda_h}
\right)^2 \right]
\right\}
\label{delta-lim}
\end{equation}

\noindent In fact, the formulas (\ref{lg})-(\ref{delta}) can also be used well beyond this range, but the
above mentioned accuracy is not guaranteed.

We also present here an approximate expression for the optimal beta-function (an accuracy
is about 10 \% in the above mentioned parameter range):

\begin{equation}
\beta_{\mathrm{opt}} \simeq 11.2 \left(\frac{I_A}{I} \right)^{1/2} \frac{\epsilon_n^{3/2}
\lambda_{\mathrm{w}}^{1/2}}
{\lambda_h  h^{1/2} K A_{JJh}} \ (1+8\delta)^{-1/3}
\label{beta}
\end{equation}

To estimate the saturation length,
one can use the result from Ref.~\cite{njp}, generalized to the case of harmonic lasing:

\begin{equation}
L_{\mathrm{sat}} \simeq 0.6 \ L_{g} \ln \left( h N_{\lambda_h} \frac{L_{g}}{\lambda_{\mathrm{w}}} \right)  \ .
\label{lsat}
\end{equation}

\noindent Here $N_{\lambda_h}$ is a number of electrons per wavelength of the considered harmonic.
For operating VUV and X-ray SASE FELs one typically has $L_{\mathrm{sat}} \simeq (10 \pm 1) \times \ L_{g}$.

Energy spread in the electron beam grows along the undulator length due
to the quantum diffusion \cite{qf-limit,qf-und}.
In this case an effective parameter $\delta$ can be introduced in order to describe
an increase in saturation length due to this effect, see Appendix B.
Let us also note that all the above presented results are reduced to those
of Ref.~\cite{des-form} for the case of the first harmonic ($h=1$).
All these results were obtained under the assumption that
beta-function is optimal (i.e. it is given by Eq.~(\ref{beta})).
However, for technical reasons this is not always the case in real machines, and it could often be
that $\beta > \beta_{\mathrm{opt}}$. In such a case the gain length can be approximated as follows:

\begin{equation}
L_{g} (\beta) \simeq L_g (\beta_{\mathrm{opt}})
\left[ 1 + \frac{(\beta-\beta_{\mathrm{opt}})^2 (1+8\delta)}{4 \beta_{\mathrm{opt}}^2} \right]^{1/6}   \ \ \ \ \ \ \ \
\mathrm{for} \ \  \beta > \beta_{\mathrm{opt}}
\label{beta-nonopt}
\end{equation}

Finally, let us note that widely used Ming Xie formulas \cite{ming-form-1,ming-form-2}
can be easily generalized to the case of harmonic lasing, see Appendix C.
Comparing two approaches to parametrization of FEL gain length, we have found that they agree reasonably well,
also for non-optimal beta-functions and well beyond the  range given by Eq.~(\ref{emit-lam-lim}).

\section{Simultaneous lasing}

In linear regime of a SASE FEL operation the fundamental frequency and harmonics grow
independently with gain lengths $L_g^{(h)}$
(here and below in this paper the superscript indicates harmonic number).
In 1D theory \cite{mcneil} the gain length of the fundamental mode is always the shortest, i.e. the fundamental
always reaches saturation first.
When analyzing parameter space in the frame of 3D theory, we realized that one can have an opposite situation in the
parameter range $2\pi \epsilon/\lambda \ll 1$, which is typical for soft X-ray beamlines of X-ray FEL facilities. This
case is discussed in Section 7. Here we consider the case $2\pi \epsilon/\lambda \simeq 1$,
or $2\pi \epsilon/\lambda \gg 1$, so
we will use the results of the previous Section. We will show
that in this regime the fundamental mode has always an advantage, i.e. its gain length is always the shortest.

Formulas of the previous Section are obtained under the condition
that beta-function is optimal for each harmonic. In this case, as one can see from (\ref{lg0}),
the gain lengths for harmonics are significantly larger than that for the fundamental mode
(this is mainly due to decrease of the wavelength, i.e.
an increase of parameter $2\pi \epsilon/\lambda$, see Appendix A for details).
For example, when the undulator parameter $K$ is large, one obtains that
$L_g^{(1)}/L_g^{(3)} \simeq 0.56$ if beta-function is optimized for each case.
However, if we consider simultaneous lasing then the beta-function
is, obviously, the same for the fundamental and for harmonics. Thus, the ratio of gain lengths depends on the choice of
beta-function.

Let us consider the case when an influence of the energy spread on FEL gain can be neglected. We find from (\ref{beta}) that
optimal beta-function for harmonics is significantly larger than that for the fundamental frequency.
If one optimizes $\beta$ for
the fundamental mode, then there is practically no lasing at harmonics. Indeed, in addition to above mentioned tendency for
optimal beta-functions, the gain lengths of harmonics will be strongly increased due to the longitudinal
velocity spread caused by too
tight focusing. In this case only nonlinear harmonic generation is possible.

If one optimizes $\beta$ for the lasing at a
harmonic, the situation can be much improved. For example, in the case of a large $K$ value,
we find from (\ref{beta}) that the optimal beta-function for the third harmonic is larger by a factor of 3.7 than that for
the fundamental frequency. Then from (\ref{lg0})
and (\ref{beta-nonopt}) one can obtain that $L_g^{(1)}/L_g^{(3)} \simeq 0.67$.

If one further increases $\beta$ such that it is much larger
than the optimal one for the considered harmonic, one can find from (\ref{lg0}), (\ref{beta}), and (\ref{beta-nonopt}) that
the ratio of gain lengths can be approximated by

\begin{equation}
\frac{L_{g}^{(1)}}{L_{g}^{(h)}} \simeq   \frac{h^{1/6} A_{JJh}}{ A_{JJ1}}  \left(
\frac{\beta_{opt}^{(h)}}{\beta_{opt}^{(1)}} \right)^{1/3} \simeq   \left( \frac{h A_{JJh}^2}{ A_{JJ1}^2} \right)^{1/3}
\ .
\label{ratio-large-beta}
\end{equation}

\noindent The considered situation corresponds to the 1D cold beam limit, and the Eq.~(\ref{ratio-large-beta})
reproduces the result of
1D model \cite{mcneil}. However, this should be considered as a coincidence since we used the
fitting formulas rather than asymptotical behavior of the exact solution of 3D theory. As an example, let us consider
again the third harmonic and large values of the undulator parameter $K$. In this case we get $L_g^{(1)}/L_g^{(3)} \simeq
0.87$, i.e. the fundamental wavelength still has an advantage, although less pronounced. Note that the inclusion
into consideration of the energy spread
effects leads always to a decrease of the considered ratio since harmonics are more sensitive to this
parameter than the fundamental mode. We conclude that in the case of the simultaneous lasing in the parameter range $2\pi
\epsilon/\lambda \simeq 1$, or $2\pi \epsilon/\lambda \gg 1$
the fundamental mode always has the shortest gain length, i.e. it saturates first.

\section{Suppression of the fundamental harmonic}

When the saturation is achieved at the fundamental frequency, the nonlinear harmonic generation occurs, i.e. the radiation
of the bunched beam at odd harmonics of the undulator \cite{flash,hg-2,hg-3,kim-1,hg-exp-1,lcls-harm}.
This radiation has a relatively low power
(for the 3rd harmonic it is on the order of a per cent of the saturated power of the fundamental wavelength),
and its relative bandwidth is about
the same as that of the fundamental \cite{harm-prst}.
Intensity of harmonics is subjected to much stronger fluctuations than that of the fundamental
frequency \cite{kim-1,harm-prst,atto-stat}.
Linear amplification of a harmonic does not proceed
due to a strong impact of the saturation at the fundamental mode on the longitudinal phase space of the electron beam.

If, however, we disrupt the lasing at the fundamental frequency such
that it stays well below
saturation, then the third harmonic lasing proceeds up to saturation resulting in a significant intensity (about 30 \% of
the saturated power of the fundamental mode in 1D limit, see Appendix D),
narrow relative bandwidth (also about 30 \% of that at the fundamental in 1D case).
In other words, the brilliance can be by two orders of magnitude higher than in the case of nonlinear harmonic generation
(for the 5th harmonic the improvement can reach three orders).
Intensity fluctuations of a harmonic are about the same as those at the fundamental wavelength of a SASE FEL since
statistics is the same.
Moreover, if the fundamental harmonic is strongly suppressed in the undulator, the users of X-ray facilities do not need filters which
are in most cases required if one uses nonlinear harmonic generation. Note that the filters suppress the fundamental wavelength
but may also partially suppress harmonics. Thus, harmonic lasing up to its saturation has decisive advantages over nonlinear
harmonic generation, so one should have good methods to disrupt the fundamental mode.

\subsection{Phase shifters}

A method to disrupt the fundamental harmonic (while keeping the lasing at the third harmonic undisturbed) was
proposed in \cite{mcneil}. The undulators for X-ray FELs consist of many segments. In case of gap-tunable undulators, phase
shifters are foreseen between the segments. If phase shifters are tuned such that the phase delay is $2\pi/3$
(or $4\pi/3$) for the fundamental, then its amplification is disrupted. At the same time the phase shift is equal to $2\pi$
for the third
harmonic, i.e. it continues to get amplified without being affected by phase shifters.
However, the simulations in \cite{mcneil}
were done for the case of a monochromatic seed,
and the results cannot be applied for a SASE FEL. The reason is that
in the latter case the amplified frequencies are defined self-consistently, i.e. there is frequency shift (red or blue)
depending on positions and magnitudes of phase kicks. This leads to a significantly weaker suppression effect. In particular, we found out (see Appendix D) that a consecutive
use of phase shifters with the same phase kicks $2\pi/3$  (as proposed in \cite{mcneil}) is inefficient,
i.e. it does not lead to a sufficiently strong suppression of the fundamental wavelength.

We propose here a modification of phase shifters method that can work in the case of a SASE FEL.
We define phase shift in the same way as it was done in \cite{mcneil} in order to make our results compatible with the
previous studies.
For example, the shift $2\pi/3$ corresponds to the
advance\footnote{In a phase shifter (like a small magnetic chicane) the beam is, obviously, delayed
with respect to electromagnetic field. One can, however, always add or subtract $2\pi$, so that the shift is kept between 0 and
$2\pi$. Therefore, a delay in the phase shifter by $2\pi/3$ corresponds to the phase shift of $4\pi/3$ according to the
definition in \cite{mcneil}, and vice versa.}
of a modulated electron beam with respect to electromagnetic field by $\lambda_1/3$.
In the following we assume that a distance between phase shifters is shorter than the field gain length of the fundamental
harmonic.
Our method of disrupting the fundamental mode can be defined as a piecewise use of phase shifters with the strength
$2\pi/3$ and $4\pi/3$.
For example, in the first part of the undulator (consisting of several segments with phase shifters between them)
we introduce phase shifts $4\pi/3$.
A red-shifted (with respect to a nominal case without phase shifters) frequency band is amplified starting up from shot
noise\footnote{A magnitude of the red shift is defined by the condition that the phase shift, $-2\pi/3$
in the considered case, is compensated by the following
phase advance in the undulator section between the phase shifters. Also sidebands (with a smaller gain) can be amplified
for which an additional phase shift in the undulator section is $2\pi$ or a multiple of it.}. In the
following second part of the undulator we use $2\pi/3$ phase shifts, so that the frequency band, amplified in the first part,
is practically excluded from the amplification process. In a realistic 3D case,
the radiation is diffracted out of the electron beam, and
the density and energy modulations within this frequency band are partially suppressed due to emittance and energy spread
while the beam is passing the second part of the undulator (although the suppression effect is often small).
Instead, a blue-shifted frequency band is amplified in the second part of the undulator, starting up from shot noise.
Then, in the third part we change back to $4\pi/3$ phase shifters, having
the residual modulations
in the electron beam and diffracted radiation from the first part as initial conditions for the red-shifted frequency band.
Then one can change to the fourth part with
$2\pi/3$ phase shifts, and so on. A more thorough optimization can also include
a part (or parts) of the undulator with zero phase shifts. As a result of these manipulations, the bandwidth of the FEL
radiation strongly increases, while the saturation is significantly delayed. The efficiency of the method strongly
depends on the ratio of the distance between phase shifters and the field gain length of the undisturbed fundamental mode.
The smaller this ratio, the stronger
suppression can be achieved after optimization of phase shifts distribution. For example, when the ratio is about 0.5, one
can relatively easy increase the "effective" gain length by a factor of 2.

An example of using this method is shown
in Fig.~\ref{xfel-62kev}, the FEL process is simulated with 3D code FAST \cite{fast} modified in order to include
harmonic lasing.
The described method also works well in 1D cold beam case (see Appendix D). The main effect here is an
efficient increase of the bandwidth due
to amplification of different sub-bands in different parts of the undulator.

We can simply generalize the method to the 5th harmonic lasing (higher harmonic numbers we do not discuss in this paper).
One can introduce a piecewise combination of
some of the phase shifts $2\pi/5$,  $4\pi/5$, $6\pi/5$, or $8\pi/5$ (for the fundamental frequency).
In this case also the third harmonic will see the disrupting
shifts, while the fifth harmonic will not be affected. If the number of phase shifters is sufficient,
the fundamental mode and the third harmonic can be strongly suppressed so that the fifth harmonic can reach saturation.

We have considered the case when a distance between phase shifters is shorter than the field gain length of
the fundamental frequency. If the distance is essentially larger, the phase shifts can still be used to delay the saturation of the
fundamental but typically the suppression effect is not sufficiently strong. However, a combination of these rare
phase shifts with intra-undulator spectral filtering can be efficient enough.

\subsection{Intra-undulator spectral filtering}

In some segmented undulator systems a number of phase shifters might not be sufficient for a required suppression of the
fundamental harmonic.
Also, some undulator systems of X-ray FELs have fixed gap, and therefore they have no phase shifters.
In this case one can either install them (if space is available) in order to have a possibility of harmonic lasing,
or to use another method, that we would like to propose here, namely an intra-undulator spectral filtering.

The idea of the method is simple: at a position in the undulator where the fundamental harmonic is in the
high-gain linear regime
(well below saturation), the electron beam trajectory deviates from a straight line, and a filter is inserted that
strongly suppresses the fundamental mode but only weakly affects the third harmonic. As a simple bending system one can
use, for example, a chicane that substitutes one of the undulator segments as it is done at LCLS for operation of the
self-seeding scheme \cite{geloni}. A possible alternative is to make a closed bump with the help of moving quadrupoles of the
undulator focusing system (quadrupoles are usually placed after each undulator segment, so that in this case
two segments are excluded from lasing). Although the main purpose of the bending system is to provide an offset for
insertion of a filter, it has to satisfy two other requirements: on the one hand
a delay of the bunch with respect to a radiation pulse
must be smaller than the bunch length; on the other hand, the $R_{56}$ (equal to the double delay)
should be sufficient for
smearing of energy and density modulations at the fundamental wavelength:
$2\pi \sigma_{\gamma}R_{56}/(\gamma \lambda_1) \gg 1$.
Both conditions can be easily satisfied simultaneously in most cases.

If the filter is efficient in suppression of the fundamental frequency (i.e. if the power is reduced to the level of the
effective power
of shot noise), after the chicane we have only the amplified radiation at the third harmonic as an input signal,
and no modulations in the beam. It means that in the second part of the undulator the fundamental mode starts up practically  from
shot noise again, so that the third harmonic can reach saturation first despite the fact that its gain length is larger.

Filters in X-ray regime can be very efficient when the ratio of photon energies is pretty large. Indeed, an attenuation in
a material depends exponentially on the product of the attenuation coefficient $\mu$ and
a material thickness $d$ (Lambert-Beer's law),
i.e. the
transmitted intensity scales as $\exp (- \mu d)$. In the region of photon energies (up to several tens of keV,
depending on material) where the photoabsorption dominates the attenuation,
the coefficient $\mu$ depends on frequency as $a \exp (-b/\omega)$. Thus, properly choosing a material and a thickness of
the filter, one can achieve the situation when losses of the third harmonic intensity are in the range of tens of per cent,
while the fundamental wavelength loses many orders of magnitude due to the double exponential suppression. An important
requirement to the filter is that it does not disturb significantly phase front of the third harmonic radiation.
It was suggested, for example, to use diamond or silicon crystals as attenuators for LCLS II \cite{lcls2} since
they are expected to not disturb phase front essentially. To provide the wavelength tunability, one can arrange a stack
of insertable filters with different thicknesses \cite{lcls2}.

The active length\footnote{A number of undulator segments contributing to lasing can be varied by different means
in order to match it to changes in wavelength and electron beam parameters.}
of the first part of the undulator is chosen such that, on the one hand, the highest possible gain is achieved;
on the other hand,
energy modulations, induced by the FEL interaction at the fundamental wavelength
(and converted to uncorrelated energy spread through the chicane), should be sufficiently small
to avoid a significant increase of gain length of the third harmonic in the second part of the undulator.
Also, there might be limitations on peak and average power load on the filter.
So, in practice in most cases
the fundamental frequency in the first part of the undulator has to stay in the exponential gain regime, and no nonlinear harmonic
generation should be expected (although in some cases the latter regime can also be considered as an option).

If one filter is not sufficient, one can use two-stage filtering. Alternatively,
a combination with phase shifters can be used. Practical application of this combination is illustrated in Section 8.
We should also note that the considered method can be used to suppress the fundamental mode and the third harmonic,
so that the fifth harmonic can lase to saturation.

In the following Section we assume that harmonic lasing occurs under the condition that lasing at
the fundamental frequency is disrupted.

\section{Harmonics versus the retuned fundamental mode}

Let us consider a harmonic lasing
and lasing {\bf at the same wavelength} with the retuned fundamental.
In other words, we reduce the wavelength of the fundamental harmonic by, for example, a factor of three
(in case of comparison with the third harmonic) by
either increasing electron energy or reducing the undulator parameter $K$. Thus, we are going to
understand if harmonic lasing can be an alternative to a standard way of reducing wavelength in X-ray FEL facilities.
Let us start with the case when we
can neglect energy spread effects ($\delta=0$). In this Section we always assume that the beta-function is tuned
to the optimum for each case.

\subsection{Reduced $K$}

In the case when the wavelength of the fundamental harmonic is adjusted by reducing parameter $K$,
the ratio of gain lengths is obtained in Appendix E, and is given by (\ref{ratio-1h-3d}):

\begin{equation}
\frac{L_{g}^{(1K)}}{L_{g}^{(h)}} =   \frac{h^{1/2} K A_{JJh}(K)}{K_{re} A_{JJ1}(K_{re})}   \ .
\label{ratio-1h-3d-text}
\end{equation}

\noindent The superscript $(1K)$ indicates that the retunig of the undulator parameter was used to reduce wavelength
of the first harmonic.
The retuned undulator parameter $K_{re}$ is given by the simple relation:

\begin{equation}
K_{re}^2 = \frac{1+K^2}{h} -1  \ .
\label{ret-k-text}
\end{equation}

\noindent Obviously, $K$ must be larger than $\sqrt{h-1}$.

It is shown in Appendix E that Eq.~(\ref{ratio-1h-3d-text}) under accepted assumptions
(no energy spread and optimal beta-function) is rather general, i.e. it is valid for any value of the parameter
$2\pi \epsilon/\lambda$. In a particular case when the approximation (\ref{lg0}) is valid,
one can also obtain (\ref{ratio-1h-3d-text}) from (\ref{lg0}) and (\ref{ret-k-text}).

For large $K$ the ratio in Eq.~(\ref{ratio-1h-3d-text}) is reduced with the help of (\ref{ret-k-text}) to a simple
form $h A_{JJh}/A_{JJ1}$, so that the gain length of the retuned fundamental mode is larger
by a factor of 1.41 (1.65) than that of the third (fifth) harmonic. In the case of a large harmonic number one can
obtain that this ratio is given by $0.94 \ h^{1/3}$ (see the asymptotic expression for $A_{JJh}$ in Section 2).

\begin{figure*}[tb]

\includegraphics[width=.8\textwidth]{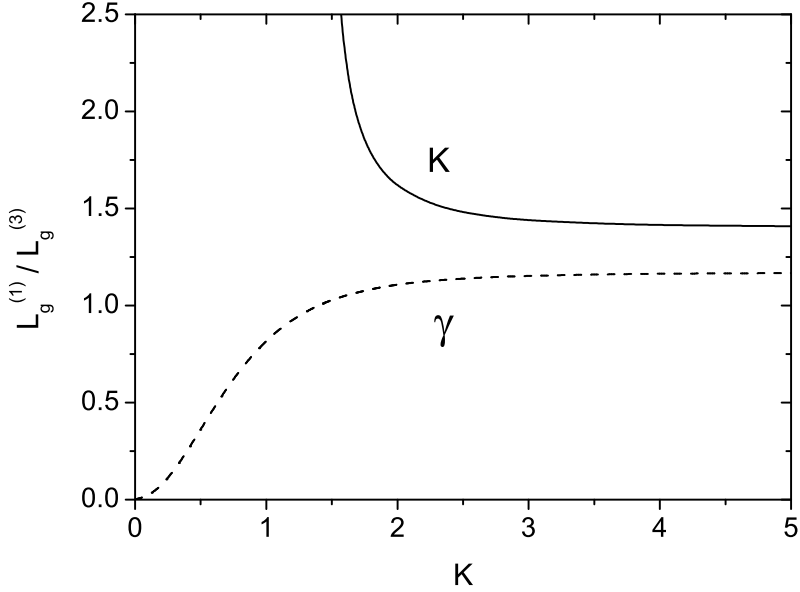}

\caption{\small Ratio of gain lengths of the retuned fundamental and the third harmonic for lasing at the same wavelength
versus rms undulator parameter $K$. The fundamental wavelength is reduced by means of reducing the undulator parameter $K$
(solid) or increasing beam energy (dash).}

\label{3d-gam-k}
\end{figure*}

For an arbitrary $K$ we plot in Fig.~\ref{3d-gam-k} the ratio of gain lengths (\ref{ratio-1h-3d-text}). It is seen that
the third harmonic always has an advantage (in case of negligible energy spread), i.e. its gain length is shorter for
any value of $K$. In Appendix E we compare this result with the result of 1D theory, and come to the conclusion that
the 3D theory actually gives more optimistic predictions in this respect, i.e. the ratio (\ref{ratio-1h-3d-text})
is always larger than the corresponding ratio of 1D theory.

\subsection{Increased beam energy}

If, instead of changing undulator parameter $K$, one changes electron energy in order to reduce fundamental wavelength by
a factor of h (i.e. one increases the energy by $\sqrt{h}$), the corresponding ratio of gain lengths at the fundamental
frequency
and at a harmonic  can be easily deduced from (\ref{lg0}):

\begin{equation}
\frac{L_{g}^{(1\gamma)}}{L_{g}^{(h)}} = \frac{h^{5/6} A_{JJh}(K)}{A_{JJ1}(K) }
\label{ratio-13-3d-gam}
\end{equation}

\noindent The superscript $(1\gamma)$ tells that the change of relativistic factor was used to reduce wavelength
of the fundamental. In Fig.~\ref{3d-gam-k} we present the ratio calculated with
(\ref{ratio-13-3d-gam}). In the case of boosting electron energy for lasing at three times reduced fundamental
wavelength, the advantage of using 3rd harmonic is not that obvious (since an increase of electron energy at the same wavelength
leads to a decrease
of the parameter $2\pi \epsilon/\lambda$ thus improving FEL properties, in general). However, even in this case, the gain length
for the third harmonic is shorter if rms value of $K$ is larger than 1.4.

\subsection{Numerical example}

Let us present a numerical example for the European XFEL \cite{euro-xfel-tdr}.
New baseline parameters \cite{tsch,winni,sy-xfel} assume operation at
different charges from 20 pC to 1 nC and three different electron energies: 10.5, 14, and 17.5 GeV.
Let us consider operation at 1 $\rm{\AA}$ with the charge 0.5 nC, peak current 5 kA, normalized emittance 0.7 $\mu$m,
and electron energy 10.5 GeV in a planar undulator with the period 4 cm. Energy spread effects are neglected here
($\delta = 0 $).
For the rms K value of 2.3 the fundamental wavelength is 3 $\rm{\AA}$, which is suppressed by using phase shifters and/or
spectral filtering\footnote{Eventually the undulator will be equipped with a chicane for operation of self-seeding
scheme \cite{geloni-xfel}.}.
Then we have third harmonic lasing at 1 $\rm{\AA}$ with the field gain length of 6.9 m according to
(\ref{lg0}) for $h=3$.
Now we change the rms K value to 1.05 so that lasing at the fundamental frequency occurs at 1 $\rm{\AA}$. In that case
we find from (\ref{lg0}) for $h=1$ that the gain length is 10.4 m,
i.e. about 50 \% larger than in the case of 3rd harmonic lasing. If, instead, we increase beam energy to 17.5 GeV and
lase at 1 $\rm{\AA}$ with $K=2.2$, the gain length is 7.9 m, i.e. it is still visibly larger than in the case of
low energy and the 3rd harmonic lasing.

\subsection{Energy spread effects}

Higher harmonics are more sensitive to the energy spread than the fundamental one \cite{kim-1,mcneil},
see the discussion below. However, a reserve in gain length in the case of no
energy spread lets harmonics be competitive with the fundamental frequency also when the energy spread effects are significant.

\begin{figure*}[tb]

\includegraphics[width=.8\textwidth]{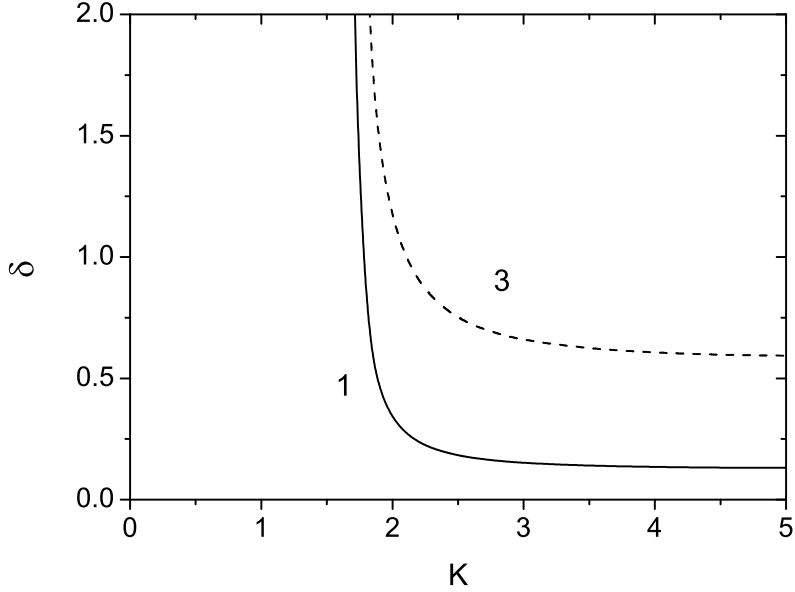}

\caption{\small Parameter $\delta$ for the retuned fundamental harmonic ($\delta^{(1)}$, solid line), and for the third harmonic
($\delta^{(3)}$, dash line), versus rms undulator parameter. The gain lengths of the third harmonic and of the retuned
fundamental harmonic are equal. Retuning is done by reducing undulator parameter.}

\label{en-sp-1-3}
\end{figure*}

To be specific, we consider the case when the fundamental wavelength is adjusted by reducing the parameter $K$.
Let the ratio of gain lengths
of a harmonic and of the fundamental mode from Eq.~(\ref{lg}) be equal to 1. Then, observing that for a given harmonic number
the ratios of parameters $\delta$ and of $L_{g0}$ are the functions of the parameter $K$ only,
one can calculate the value of $\delta$, for which the gain lengths are equal, as a function of $K$. In Fig.~\ref{en-sp-1-3}
we plot such a dependence for the case of the third harmonic lasing.
Continuing the numerical example of the previous Section for 1 $\rm{\AA}$ operation at 10.5 GeV,
we find that the gain lengths of the fundamental mode and of the third harmonic become equal to 12.5 m
when the energy spread is 2.8 MeV. For smaller values of the energy spread the third harmonic has shorter gain length
than the fundamental one.

In the case of going to higher beam energy for reducing the fundamental wavelength, the margins for energy spread are
reduced in the case of harmonic lasing.
Still, acceptable values of the energy spread can be relatively large.
In particular, in the considered numerical example (with 17.5 GeV for lasing at the fundamental frequency)
the third harmonic lasing at 10.5 GeV has a shorter gain length when the energy spread is below 1.3 MeV.

\subsection{Fifth and higher harmonics}

To let the fifth harmonic lase to saturation, one has to suppress lasing at the fundamental frequency and at the third harmonic.
This can be done as discussed above in this paper: by using either a special set of phase shifters or intra-undulator
spectral filtering or a combination of these two methods. Higher harmonic numbers we do not consider in this paper, but
the tendency can be summarized as follows. In the case of a sufficiently large $K$ value the gain length gets shorter
with an increase of harmonic number (as discussed above) if there is no energy spread.
However, the sensitivity to the energy spread also
strongly increases, so that at some point there is a cutoff. From practical point of view, one has to consider limitations
due to undulator phase errors, undulator wakefields etc. These issues are discussed in Section 8.
The use of the fifth harmonic can still be considered in some cases as quite realistic.

For illustration, let us continue the numerical example for the European XFEL with the energy of 10.5 GeV. In order
to lase at
1 $\rm{\AA}$ with the fifth harmonic, one has to increase $K$ to 3.1, so that the fundamental wavelength is 5 $\rm{\AA}$.
In case of negligible energy spread the field gain length for the fifth harmonic is 5.8 m (to be compared with 10.4 m
for the retuned fundamental harmonic). When the energy spread is 2.3 MeV, the gain length is the same in both cases and
equals 11.8 m.

\subsection{Discussion on 3D and warm beam effects}

An obvious disadvantage of higher harmonics in comparison with the fundamental harmonic is a weaker coupling between
the electron current and the electromagnetic field, described by the coupling factor $A_{JJh}$.
When one considers lasing at the same wavelength with a harmonic and a retuned fundamental, the main advantage of
the harmonic is connected with a higher mobility of particles which allows them to get bunched easier. Indeed, the derivative
of the longitudinal dispersion $R_{56}' = (1+K^2)/\gamma^2$ is proportional to a harmonic number, no matter if one
changes $K$ or beam energy in order to adjust wavelength of the fundamental harmonic. As a net effect, in a simple 1D model
harmonics have an advantage as
soon as $K$ is sufficiently large so that the $A_{JJh}$ is not too small. In a particular case of retuning $K$ in order
to adjust wavelength of the fundamental, harmonics have always an advantage \cite{mcneil}
since the retuned $K$ for the fundamental frequency gets too small before $A_{JJh}$ for a harmonic drops.

An inclusion of such 3D effects as diffraction of radiation and transverse motion of particles does not change the ratio
of gain lengths significantly since we are considering the same wavelength. But a difference with 1D model is also
connected with the possibility
to optimize beta-function. If $\beta$ is too large, the current density is too small and FEL gain is weak;
if $\beta$ is too small, the longitudinal velocity spread due to emittance suppresses the FEL gain.
As a result, there is always an optimum.
Note that an effect of the longitudinal velocity spread due to emittance has nothing to do with the above mentioned mobility,
characterized by longitudinal dispersion (which is important for energy spread effect, see below).
When higher harmonics have an advantage in zero order, as explained above, they are somewhat less sensitive
to this velocity spread than the retuned fundamental harmonic, and the optimal $\beta$ is smaller for harmonics. Thus,
due to tighter focusing one gets smaller beam size and therefore, an additional advantage over the fundamental frequency.
This explains the fact that the 3D model (with optimized beta-function) gives more advantage in growth rate to
harmonics than the 1D model does (see Appendix E).

Finally, harmonics are more sensitive to the energy spread due to a larger $R_{56}'$. However,
when they have a significant advantage in the case of negligible energy spread, a relatively large correction to the
gain length due to a finite energy spread can be tolerated, as one can see from numerical examples. Note that $\beta$
is adjusted depending on the energy spread, what cannot be done in a simple 1D model.

Concluding this Section, we can state that harmonic lasing is
especially attractive in the case of gap-tunable undulator when lasing at the shortest wavelength is achieved with
the opened gap, i.e. when $K$ is reduced. In fact, the highest photon energy of an X-ray FEL facility, at which
saturation occurs, in case of
the 3rd harmonic lasing can be typically increased by 30-100 \%.
On the other hand, harmonic lasing at a reduced electron energy is a possible solution for a compact and relatively
cheap X-ray FEL facility.

\section{Properties at saturation and a possible increase of brilliance}

FEL properties at saturation can be calculated with the help of a numerical simulation code (for 1-D simulations
see Ref.~\cite{mcneil} and Appendix D). Here we present a qualitative consideration for the case when
the energy spread effect is a relatively weak correction
to the FEL operation ($\delta \ll 1$), and the tuning to the same wavelength is achieved by changing parameter K.
A simple estimate ("effective" parameter $\rho$ \cite{bon-rho} is reduced depending on harmonic number)
suggests that in the case of harmonic lasing, both the saturation power and the
bandwidth are reduced by the same factor. Degree of transverse coherence is about the same
for a harmonic and for the fundamental mode since this quantity is mainly defined \cite{tr-coh-1,tr-coh-2} by
the parameter $2 \pi \epsilon /\lambda$, which is the same in the considered case.
Thus, the brilliance (a figure of merit for performance of X-ray FELs), depending on the ratio of peak power to bandwidth,
remains about the same. In other words, use of harmonic lasing instead of lasing at the fundamental frequency
is equivalent to a mild monochromatization of the X-ray beam.

Here we propose a simple method of brilliance improvement.
In a gap-tunable undulator one can combine a high power and a narrow bandwidth.
A possible trick is to use harmonic lasing in the exponential gain regime in the first part of the undulator,
making sure that the fundamental frequency is well below saturation (two options can be considered:
with and without disruption of
the fundamental by phase shifters, depending on the ratio of gain lengths).
In the second part of the undulator the value of K is reduced such that now the fundamental mode is resonant to the wavelength,
previously amplified as the third harmonic.
The amplification
process proceeds in the fundamental mode up to saturation. In this case the bandwidth is defined by the harmonic lasing
(i.e. it is reduced by a significant factor depending on harmonic number) but the saturation power is still as high as in the
reference case of lasing at the fundamental, i.e. brilliance increases. Important is that this option
does not require extra undulator length.

\section{Simultaneous lasing in the case of a thin electron beam}

For a typical operating range of hard X-ray FELs the condition $2\pi \epsilon/\lambda \simeq 1$ is usually a design goal
for the shortest wavelength.
In the case of the simultaneous lasing the fundamental mode has shorter gain length than harmonics,
as it was shown
above in this paper. However, if the same electron beam is supposed to drive an FEL in a soft X-ray beamline,
the regime with $2\pi \epsilon/\lambda \ll 1$ is automatically
achieved. Here we present a detailed study of this regime. In this Section we assume
that beta-function is sufficiently large, $\beta \gg L_g^{(h)}$. In this case we can use the model of parallel beam (no
betatron oscillations), and can also neglect an influence of longitudinal velocity spread due to emittance on FEL process.
If in addition the energy spread is negligibly small, then the normalized FEL growth rate at the fundamental frequency is described
by the only
dimensionless parameter, namely the diffraction parameter B \cite{book}, see appendix A.
The generalized diffraction parameter
$\tilde{B}$, that can be used for harmonics, is also introduced in Appendix A, see (\ref{diff-new}). We rewrite it here as
follows:

\begin{equation}
\tilde{B} = 2 \epsilon \beta \tilde{\Gamma} \omega_h /c \ ,
\label{diff-new-1}
\end{equation}

\noindent where $\omega_h = 2\pi c/\lambda_h$ and $\tilde{\Gamma}$ is the gain factor that also depends on harmonic number:

\begin{equation}
\tilde{\Gamma} =
\left( \frac{A_{\mathrm{JJh}}^2 I\omega_h ^{2} K^2(1+K^2)}{I_{\mathrm{A}}c^{2}\gamma^5} \right)^{1/2}
\label{gain-param-1}
\end{equation}

The gain length of a harmonic is defined by the universal function of $\tilde{B}$:

\begin{equation}
L_{g}^{(h)} = [ \tilde{\Gamma} f_1 (\tilde{B}) ]^{-1}
\label{gain-l-thin}
\end{equation}

The function $f_1(\tilde{B})$ can be calculated from the general eigenvalue equation (\ref{eq:hank-trans-1}).
However, within the parallel beam model,
accepted in this Section, the eigenvalue equation can be significantly simplified.
We use here the solution of the equation
presented in \cite{book,opt-com-93} for the Gaussian transverse distribution of current density (see Fig.~4.52 of
Ref.~\cite{book}). In the parameter range, that is the most interesting for our purpose, we can approximate the function
$f_1(\tilde{B})$ as follows:

\begin{equation}
f_1(\tilde{B}) \simeq 0.66 - 0.37 \log_{10}(\tilde{B})  \ \ \ \ \ \ \ \ \ \ \ for \ \ \tilde{B} < 3 \ .
\label{fd}
\end{equation}

Using the superscript $(h)$ to indicate the harmonic number
for the diffraction parameter and the gain factor, we can see that

\begin{equation}
\frac{\tilde{B}^{(h)}}{\tilde{B}^{(1)}} = \frac{h \tilde{\Gamma}^{(h)}}{\tilde{\Gamma}^{(1)}} = \frac{h^2 A_{JJh}}{A_{JJ1}} \ .
\label{b-h}
\end{equation}

According to (\ref{gain-l-thin}) and (\ref{gain-param-1}), the ratio of gain lengths can be presented as follows:

\begin{equation}
\frac{L_{g}^{(1)}}{L_{g}^{(h)}} = \frac{h A_{JJh}}{A_{JJ1}} \
\frac{ f_1 (\tilde{B}^{(h)})}{f_1 (\tilde{B}^{(1)})}
\label{gain-l-h-thin}
\end{equation}

\begin{figure*}[tb]

\includegraphics[width=.8\textwidth]{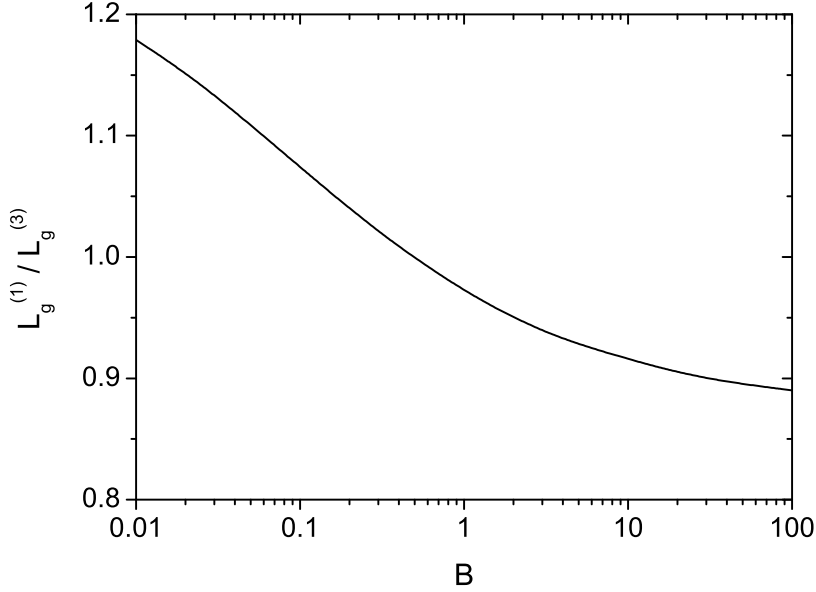}

\caption{\small
Ratio of gain lengths for lasing at the fundamental wavelength and at the third harmonic versus diffraction
parameter of the fundamental wavelength for large values of the undulator parameter K.
}
\label{thin-1-3}
\end{figure*}

One can easily observe from (\ref{b-h}) and (\ref{gain-l-h-thin})
that for a given value of diffraction parameter for the fundamental frequency, $B = \tilde{B}^{(1)}$, this ratio
depends only on the parameter $K$ for a considered harmonic. If $K$ is sufficiently large (see Fig.~1),
one can obtain a universal dependence which is presented in Fig.~\ref{thin-1-3} for the case of the third harmonic.
For large values of the diffraction parameter (wide electron beam limit)
one can use an asymptotic expression for the growth rate \cite{book}, so that
the function $f_1$ is proportional to $(\tilde{B}^{(h)})^{-1/3}$. In this case one obtains the result of
1D theory \cite{mcneil}:

\begin{displaymath}
\frac{L_{g}^{(1)}}{L_{g}^{(h)}} \simeq  \left( \frac{h A_{JJh}^2}{ A_{JJ1}^2} \right)^{1/3}
\ .
\end{displaymath}

In the case of the third harmonic and large $K$ this ratio is equal to 0.87. One can see that the curve in
Fig.~\ref{thin-1-3} slowly approaches this value when $B$ is large. So, in the limit of wide electron beam, corresponding
to 1D model, the fundamental frequency has shorter gain length than harmonics.

In the limit of small diffraction parameter (thin
electron beam) we wave the opposite situation, as one can see from Fig.~\ref{thin-1-3}.
When diffraction parameter is
smaller than 0.4, the gain length of the fundamental frequency is larger than that of the third harmonic for large values of $K$.
A similar dependence can be calculated for the fifth harmonic, in this case
the gain length of the fundamental harmonic is larger than that of the fifth harmonic (for a sufficiently large $K$) when $B<0.28$.
Moreover, the fifth harmonic grows faster than the third one when $B<0.15$ and $K$ is large. In fact, if the diffraction
parameter for the fundamental harmonic is about $0.1$ or less, there might a number of amplified harmonics
with similar growth rates. We should note that this number can be reduced when the energy spread is included
into consideration.

\begin{figure*}[tb]

\includegraphics[width=.8\textwidth]{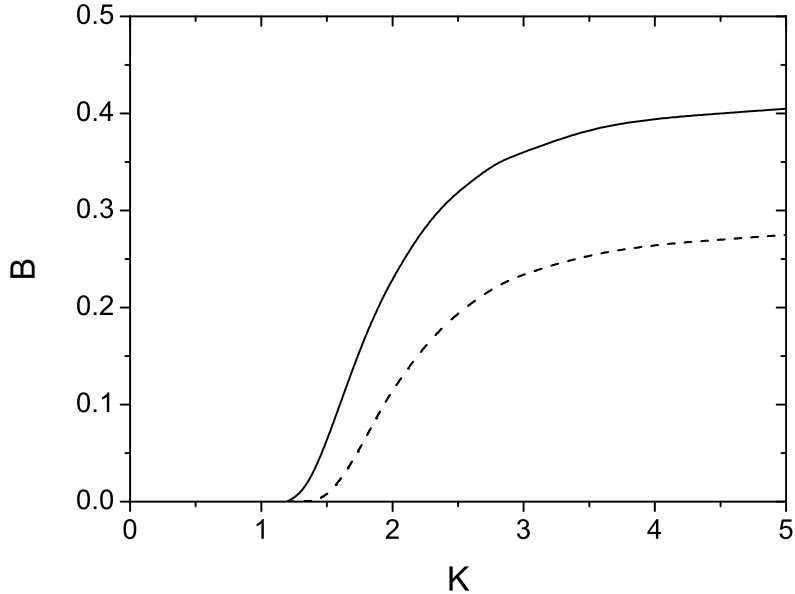}

\caption{\small
Diffraction parameter of the fundamental wavelength, for which the third (solid) and the fifth (dash) harmonics
have the same gain length as the fundamental, versus the rms undulator parameter K. Below these curves harmonics
have shorter gain lengths than the fundamental frequency.
}
\label{thin-b-k}
\end{figure*}

To find out how the value of $B$, at which the harmonics have the same gain length as the fundamental, depends on the
undulator parameter $K$, one can use the Eqs.~(\ref{fd})-(\ref{gain-l-h-thin}). We present the results for the third and
the fifth harmonics in Fig.~\ref{thin-b-k}. The areas below the curves in Fig.~\ref{thin-b-k} correspond
to the case when corresponding harmonics grow faster than the fundamental frequency. We should stress that the condition
$2\pi \epsilon/\lambda \ll 1$ is necessary but not sufficient for reaching this regime.

Let us discuss why the effect, considered in this Section, can only take place in the frame of 3D theory and
in the limit of a thin beam. In 1D theory the gain factor (inversely proportional to the gain length)
scales as $(A_{JJh}^2 \omega_h)^{1/3}$, if we keep only parameters that depend on harmonic number. The frequency here comes
from the dynamical part of the problem, it reflects the fact that the beam gets bunched easier at higher frequencies.
As for the electrodynamic part of the problem, the amplitude of the radiation field of charged planes does not depend
on frequency. Since the product $A_{JJh}^2 h$ decreases with harmonic number for any $K$, gain length of harmonics is
always larger than that of the fundamental frequency. Concerning the 3D theory, the solution of the paraxial wave equation shows that
on-axis field amplitude is proportional to the frequency. So, both dynamical and electrodynamic parts
contribute to the solution of the self-consistent problem with $\omega_h$.
That is why in the gain factor in Eq.~(\ref{gain-param-1}) we
have squared frequency $(A_{JJh}^2 \omega_h^2)^{1/2}$, i.e. it depends on harmonic number via the product
$A_{JJh}^2 h^2$ which can increase with harmonic number if $K$ is sufficiently large. Since in the case
of a thin electron beam the function $f_1$ depends
only weakly, in fact logarithmically, on the diffraction parameter (which is larger for harmonics),
harmonics can grow faster than the
fundamental frequency in some range of parameters $B$ and $K$, as it is illustrated in Fig.~\ref{thin-b-k}.

So far we have discussed an exponential gain regime and did not consider an initial-value problem.
In the simulations one can observe that the fundamental dominates
saturation regime even if its gain length is slightly longer than that of harmonics. First, it has a higher effective
start-up power due to a larger factor $A_{JJ}$. Second, in nonlinear regime the longitudinal phase space of
the electron beam is affected stronger by the fundamental frequency. As a result, saturation power of harmonics in the
case $B \simeq 0.1$ is weaker\footnote{The third harmonic saturates earlier than the fundamental, and
at a full expected power when diffraction parameter is on the order of 0.01.}than it would have been in the absence
of the fundamental frequency (but still much higher than
in the case of nonlinear harmonic generation). The bandwidth at saturation is inversely proportional to harmonic number
(contrary to the case of nonlinear
harmonic generation).

\begin{figure*}[tb]

\includegraphics[width=.8\textwidth]{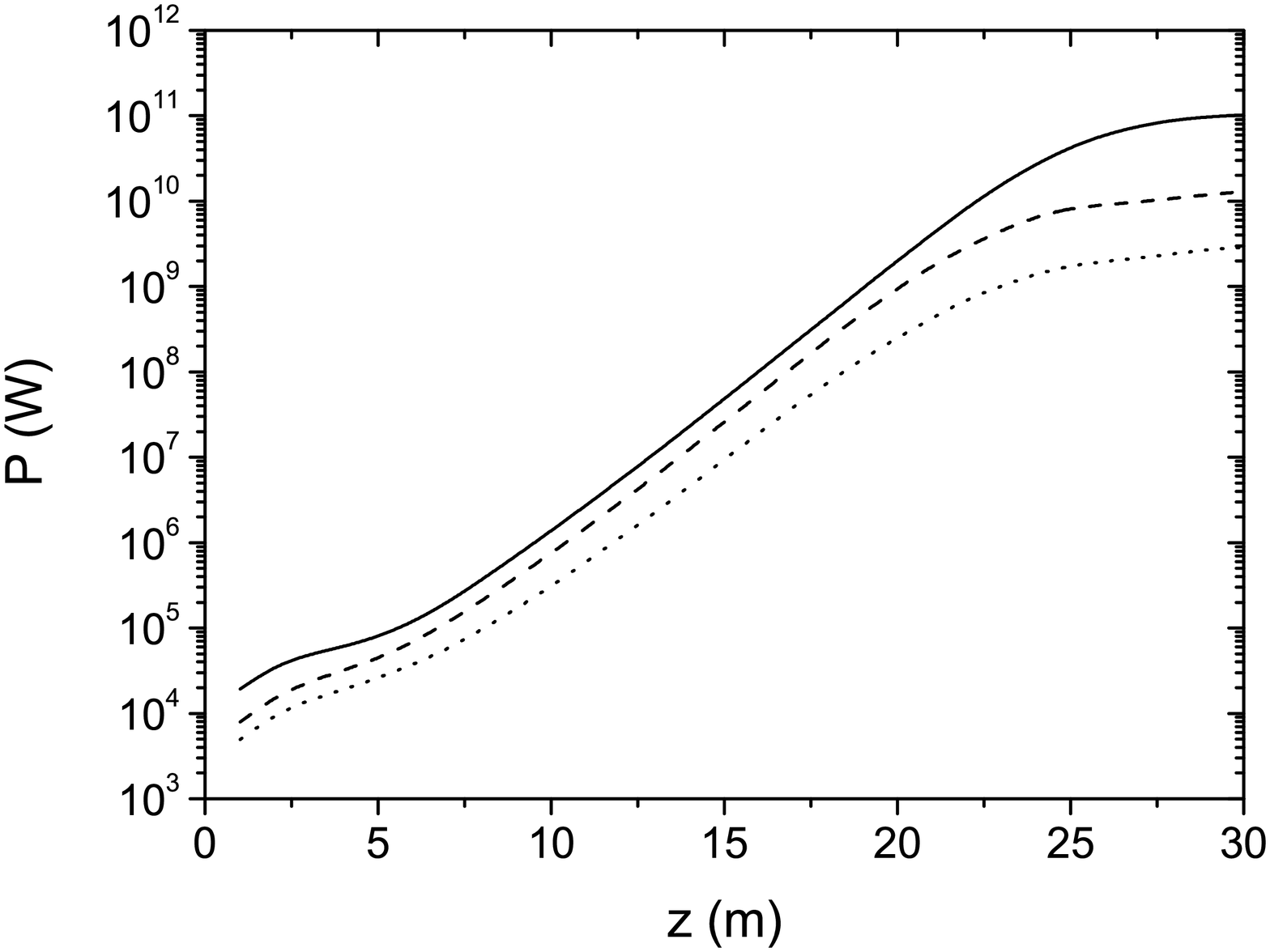}

\caption{\small
An example for the European XFEL.
Averaged peak power for the fundamental harmonic (solid), the third harmonic (dash), and the fifth harmonic (dot) versus
undulator length for SASE3 operating at 4.5 nm. Parameters are in the text. Simulations were performed with the code FAST.
}
\label{sase3}
\end{figure*}

Let us present a numerical example for the European XFEL. An electron beam with the energy of 10.5 GeV lases in
SASE3 undulator (which is placed behind the hard X-ray undulator SASE1)
with the period 6.8 cm and the rms undulator parameter 7.4 at the fundamental wavelength 4.5 nm.
We consider electron bunches with  the charge of 100 pC: the peak current is 5 kA,
averaged normalized slice emittance is 0.3 $\mu$m from start-to-end
simulations \cite{winni}, and slice energy spread\footnote{Energy spread due to the
quantum diffusion \cite{qf-und} in SASE1 and the active part
of SASE3 undulators is added quadratically to the value obtained in start-to-end similations \cite{winni}.
It is assumed that SASE3 operates in "fresh bunch" mode, i.e. there is no lasing to saturation in SASE1.} is 1 MeV.
For the beta-function of 15 m we obtain from (\ref{diff-new-1}) that the diffraction parameter for the fundamental wavelength
is 0.3, so that a simplified model, considered in this Section, suggests
that the third harmonic can grow faster than the fundamental. However, harmonics are more sensitive to the energy spread than
the fundamental frequency, therefore we use a general eigenvalue equation (\ref{eq:hank-trans-1}) that includes all the important effects.
We find that the field gain length is 2.44 m for the fundamental harmonic, 2.42 m for the third harmonic, and 2.52 m for the fifth one.
In Fig.~\ref{sase3} we present the results of numerical simulations.
Even though the saturation power of harmonics is lower than it would have been in the absence of the fundamental, it is still
by an order of magnitude higher than that expected from nonlinear harmonic generation \cite{sy-xfel}. The saturation
power of the third (fifth) harmonic is 12\% (3\%) of the saturation power of the fundamental frequency.
Thus, accurate calculation of harmonic lasing is necessary for planning of user experiments and X-ray beam transport.

Note that the method of brilliance improvement, described in the previous Section,
is especially attractive in the considered regime. Indeed, one can, in principle, use a high harmonic number so that
the bandwidth reduction can be significant. Another useful application is the simultaneous lasing at the
fundamental wavelength and at the third harmonic with comparable intensities that can be used in
jitter-free pump-probe experiments making use of a split-and-delay stage \cite{feldhaus}. For such an experiment
one can, in principle, manipulate relative intensities with the help of phase shifters.

On the other hand, a high-intensity harmonic
radiation can disturb some experiments, or may lead to an excessive power load on mirrors of X-ray transport.
In this
case the harmonics can be suppressed by different means. For example, one can increase the energy spread with the help
of a laser heater \cite{lsc,lcls-heater,lcls-heater-op} which is going to be a part of the
standard design of an X-ray FEL accelerator complex.
In the above presented example, an increase of the energy spread up to 5 MeV would strongly suppress harmonic lasing,
so that one would get an intensity level expected from nonlinear harmonic generation. Another method is the
use of phase shifters, but now aiming at suppression of harmonics. In this case the phase shifts for the
fundamental frequency could be below 1 rad while for harmonics they are
$h$ times larger, i.e. the suppression effect is stronger.
Other options are an increase of the beta-function (what leads to an increase of the diffraction parameter)
or the application of linear undulator taper \cite{stup,chirp-tap} that
would have stronger effect on the amplification of harmonics.

\section{Practical applications of harmonic lasing in X-ray FELs}

\subsection{LCLS: intra-undulator spectral filtering and phase shifters}

Linac Coherent Light Source (LCLS) is the first hard X-ray free electron laser \cite{lcls}.
Due to the limited electron energy and
fixed-gap undulator, the facility can presently cover photon energy range up to 10 keV. Nonlinear harmonic generation
was studied in \cite{lcls-harm} with the third harmonic at 25 keV, and the second harmonic afterburner
\cite{nuhn-after} operation was demonstrated at 18 keV, but the intensity was relatively low in both cases.
Here we present a numerical example for
third harmonic lasing at LCLS up to the photon energy of 25-30 keV
with a significant power and a relatively narrow intrinsic bandwidth.

LCLS undulator \cite{nuhn} consists of 33 identical 3.4-m-long segments, undulator period is 3 cm, and the peak
undulator parameter is 3.5 (rms value of $K$ is 2.5). The 16th segment is replaced with a chicane for operation of
the self-seeding scheme \cite{geloni}. When this scheme is operated, a crystal monochromator is inserted on-axis
while the electron beam goes through the chicane thus by-passing the crystal. We notice that a simple add-on to this
setup, namely an insertable filter, would allow the use of the intra-undulator spectral filtering method described
in Section 4.2.
As a possible realization of the filter we propose here a silicon crystal\footnote{Diamond can be considered as an
alternative} that is not supposed to spoil phase front \cite{lcls2} of
the third harmonic radiation while attenuating the fundamental harmonic by orders of magnitude.
A thickness of the crystal is defined by a required attenuation factor and an expected photon energy range. As an example
we consider here the thickness of 600 $\mu$m and third harmonic lasing at 25 keV.
Attenuation length at 8.3 keV is $\mu^{-1} = 73 \ \mu$m, and at 25 keV it is
$\mu^{-1} = 1.85$ mm \cite{mucal}, so that the corresponding transmission factors are $2.7 \times 10^{-4}$ and 0.72.
With a given thickness of the crystal the scheme would work well
in the range 20-30 keV, and for lower photon energies of the third harmonic a thinner crystal would be needed.

In the considered parameter range the spectral filtering method alone is not sufficient, therefore we suggest to combine
it with the phase shifters method. We propose to install phase shifters with the shift
$4\pi/3$ (the definition of Ref.~\cite{mcneil} is used here, see Section 4.1) after undulator segments 1-5 and 17-22,
and with the shift $2\pi/3$ after segments 6-10 and 23-28. As a possible space-saving technical solution one can consider
insertable permanent-magnet phase shifters with a length of a few centimeters and a fixed phase shift. Of course,
if space allows, the tunable (electromagnetic or permanent-magnet) phase shifters would be more flexible. Note also that
phase shifters without spectral filtering might not be sufficient for a sure suppression of the fundamental harmonic.

Let us consider a specific parameter set for third harmonic lasing at 0.5 $\rm{\AA}$ (photon energy 25 keV).
The electron beam
parameters are as follows: energy is 13.6 GeV (the fundamental wavelength is 1.5 $\rm{\AA}$), peak current is 3 kA,
normalized slice emittance is 0.3 $\mu$m, uncorrelated energy spread is 1.4 MeV. The beta-function in the undulator is 30 m.
Parameters of the chicane are chosen as described in Section 4.2, the smallest possible delay (given by either the required
beam offset or minimum $R_{56}$ for smearing of beam modulations at the fundamental wavelength)
would define the shortest
electron bunch that can be used for operation of this scheme. In our simulations we do not consider a specific bunch length,
so that our result is the peak power of the third harmonic radiation in the part of the pulse that overlapped with the
electron beam after the chicane. One should also notice that an easy control of the third harmonic pulse duration is possible
by changing the delay.

\begin{figure*}[tb]

\includegraphics[width=.8\textwidth]{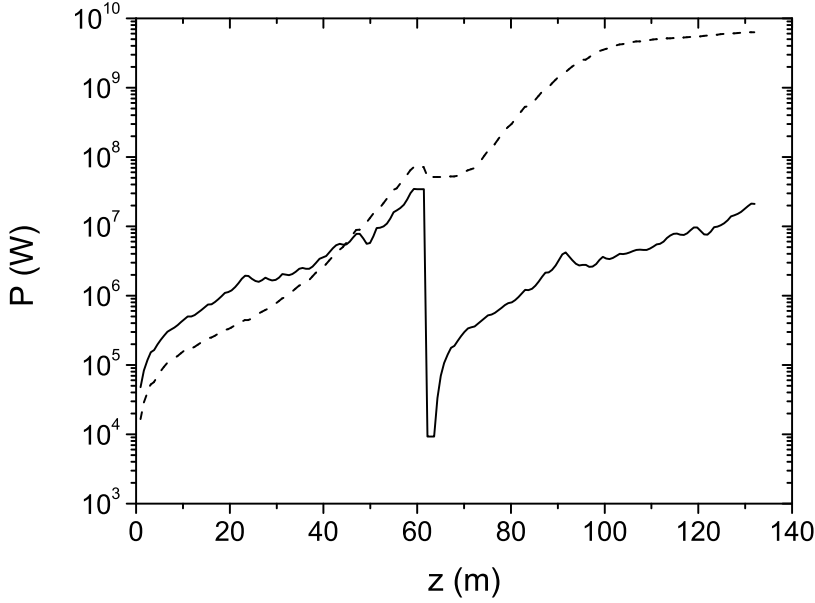}

\caption{\small
Averaged peak power for the fundamental harmonic (solid) and the third harmonic (dash) versus
geometrical length of the LCLS undulator (including breaks).
The wavelength of the third harmonic is 0.5 $\rm{\AA}$ (photon energy 25 keV).
Beam and undulator parameters
are in the text. The fundamental is disrupted with the help of the spectral filter (see the text)
and of the phase shifters.
The phase shifts are $4\pi/3$ after segments 1-5 and 17-22, and $2\pi/3$ after segments 6-10 and 23-28.
Simulations were performed with the code FAST.
}
\label{lcls-25kev}
\end{figure*}

We performed simulations with the code FAST \cite{fast}, the results are presented in Fig.~\ref{lcls-25kev}. The averaged peak power
of the third harmonic radiation is 6 GW, and an intrinsic bandwidth is $3 \times 10^{-4}$ (FWHM). The power
incident on the crystal is in the range of tens of megawatts, and should not
be problematic from the point of view of peak and average power load.
Note that the saturation of the third harmonic lasing is achieved after 28th segment, so that there is
a sufficient contingency for given wavelength and beam parameters. It means, in particular, that the saturation at 30 keV
could be in reach, or the saturation at 25 keV with a larger emittance is possible. We should also note that a
reduction of the beta-function would increase the contingency. If one considers the scheme for operation in the range
10-20 keV, it would work with a significantly loosened requirements on the electron beam quality.

As a quick test \cite{ratner} of harmonic lasing at LCLS one can consider operation with the filter only
(without phase shifters), making use of
nonlinear generation of the third harmonic in the first part of the undulator if the fundamental harmonic enters nonlinear
regime there. The main issues, that were discussed in Section 4.2, are high power load on the filter and an increase of
energy spread in the beam. However, the last issue might be partially tolerated. Indeed, in a SASE FEL
the radiation intensity and beam modulations in energy and density consist of random spikes that have a typical
duration of FEL coherence time. Thus, energy spread after the chicane is modulated on the same time scale.
One can have the situation when some of the third
harmonic intensity spikes overlap after the cicane with unspoiled parts of the electron beam,
and are amplified in the second part of the
undulator without gain suppression due to a large energy spread (however, the slippage effects in the second part
must be considered). In principle, these spikes can reach saturation in the second part at a high power level
before they are caught up by the fundamental harmonic.

\subsection{European XFEL: free option}

\begin{figure*}[tb]

\includegraphics[width=.8\textwidth]{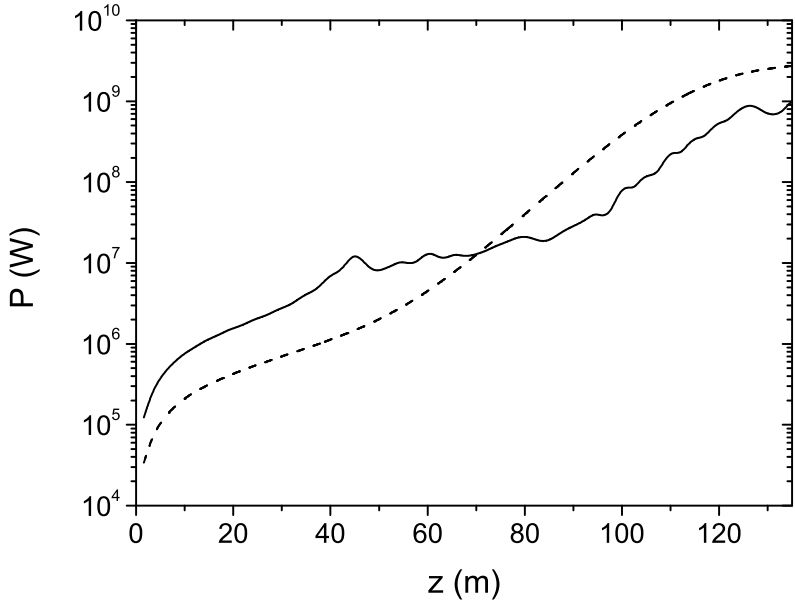}

\caption{\small
An example for the European XFEL.
Averaged peak power for the fundamental harmonic (solid) and the third harmonic (dash) versus
magnetic length of SASE1 undulator. The wavelength of the third harmonic is 0.2 $\rm{\AA}$ (photon energy 62 keV).
The fundamental is disrupted with the help of phase shifters installed after 5 m long undulator segments.
The phase shifts are $4\pi/3$ after segments 1-8 and 21-26, and $2\pi/3$ after segments 9-16.
Simulations were performed with the code FAST.
}
\label{xfel-62kev}
\end{figure*}

The gap-tunable hard X-ray undulators SASE1 and SASE2 of the European XFEL
consist of 35 segments each \cite{tsch}, the length of a
segment is 5 m, the undulator period is 4 cm. The phase shifters are installed between the segments,
so that the number of the shifters is big.
This means that, at least in some cases, the phase shifter method alone might be sufficient for suppression of the
fundamental harmonic. As an example we consider
the third harmnonic lasing at 0.2 $\rm{\AA}$ (photon energy 62 keV) by the electron beam with the energy of 17.5 GeV and
the charge of 100 pC, slice parameters are the same as those given in Section 7, beta-function is 60 m, the rms undulator
parameter is 1.6. Note that
the considered wavelength cannot be reached by lasing at the fundamental harmonic because the undulator parameter
is too small in this case.
The results of numerical simulations are
presented in Fig.~\ref{xfel-62kev}. Indeed, one can disrupt the fundamental harmonic and let the third harmonic saturate.
The averaged peak power is 3 GW, and the bandwidth is $2\times 10^{-4}$ (FWHM).
One can still notice that a stronger
suppression of the fundamental would be desirable, so that the spectral filtering method would improve operation
of the facility in such a regime. Eventually, the self-seeding scheme \cite{geloni-xfel} will be implemented at the
European XFEL, then it is also worth to install a  filter. Another option is a closed bump (made by movable
quadrupoles between the segments). Such a bump involves two segments with an insertable filter installed between them.
We should note that
if we consider a 20 pC electron bunch with slice parameters from start-to-end simulations \cite{winni}, the
third harmonic lasing to saturation can be extended to photon energies up to 100 keV.

The extension of the photon energy range is the main application of harmonic lasing at the European XFEL. A related
application is a more flexible operation of the facility, for example operation of SASE1 at a longest possible wavelength
(with the closed gap), and of SASE2 at a very short wavelength in the regime of harmonic lasing. Note, however that
the considered options might be significantly limited if the longitudinal space charge driven
microbunching instability \cite{fel-bc,lsc} will have to be cured
by the laser heater \cite{lsc,lcls-heater,lcls-heater-op} that increases the energy spread.

Another attractive option that one can consider in the case of the European XFEL is a reduction of the
bandwidth by going
to harmonic lasing instead of lasing in the fundamental mode. If one combines them as described in Section 6, this will
happen without reduction of power, i.e. the brilliance will increase. Although this increase is essentially smaller than in
the case of application of seeding and self-seeding schemes,
the method of combined lasing does not require extra undulator length,
is not restricted by a finite wavelength interval, and is completely based on a baseline design. For many
experiments, however, such a mild reduction of the bandwidth (to the level of few $10^{-4}$) would be desirable. The
detailed numerical simulations of combined lasing will be presented elsewhere.

As it was discussed in Section 7, the simultaneous lasing at the fundamental and the third harmonics with comparable
intensities for jitter-free pump-probe intensities can easily be done in SASE3 beamline. Now we can also notice
that the same holds in the case of the two hard X-ray beamlines, if one uses phase shifters to suppress the
fundamental mode as it was illustrated above. One can easily control relative intensities of the
fundamental and the third harmonics by changing phase shifters.

\subsection{Other facilities and applications}

Similar improvements can be realized at other X-ray FEL facilities like SACLA, FLASH II, LCLS II etc.
In particular, FLASH II can cover the whole water window with the help of the third harmonic lasing.
According to our estimates, one would need to combine phase shifters with intra-undulator spectral filtering, so that
some modifications in the undulator beamline would be necessary.

We also suggest that even if fixed-gap undulators are supposed to be used in some X-ray FEL facilities, phase shifters
should be incorporated into an undulator design such that the option of harmonic lasing is enabled.

We should also note that electromagnetic phase shifters can have an advantage in comparison with permanent-magnet ones.
Indeed, the electromagnetic phase shifters can be switched on and off between electron pulses (or macropulses)
so that one can switch between two colors and then to deliver them to different experiments.

One more application of the harmonic lasing is the operation of compact X-ray FEL facilities using relatively low
energy of electron beams. In order to reach short wavelength they rely on the technology of short-period small-gap
in-vacuum undulators \cite{sacla,swiss}. As an alternative one could consider a more standard and robust undulator technology by
going to larger period undulators with larger gaps, and lasing at the third (or even the fifth) harmonic. We estimated,
in particular, with the help of formulas of Section 2 that it could be an interesting option
for the Swiss FEL \cite{swiss}.

\subsection{Possible technical issues}

One can identify possible issues using the fact that the harmonic lasing is more narrow-band process than lasing
at the fundamental wavelength (parameter $\rho$ \cite{bon-rho} is smaller). In particular, harmonic lasing is
more sensitive to the undulator phase errors, undulator wakefields etc.

However, undulators for X-ray FELs are usually designed and built with a sufficient safety margin in terms of
length and quality. For example, LCLS undulator has
phase shake about 1-2 deg for the fundamental wavelength \cite{nuhn},
which means 3-5 deg for the third harmonic. This is still too small phase shake to have a significant impact on
the third harmonic lasing.

The wakefields in the undulator can have an impact on the amplification process in X-ray FELs \cite{stup},
a relevant parameter is a relative energy change per gain length, divided by $\rho$. However,
they can be compensated by using undulator
taper as it was successfully demonstrated at LCLS. The harmonic lasing is more sensitive to wakefields, however
a proper optimization of the taper would allow to diminish this effect in many practical situations, especially for
low charge scenarios.

We have shown in the paper that the application of harmonic lasing allows to generate FEL radiation with intrinsically
narrow bandwidth. However, from practical point of view it is important that the energy chirp in the electron beam and
energy stability are smaller than a half of that bandwidth.
For example, in the case of the European XFEL the chirp can be kept under control and the
energy stability can be provided at the level of $10^{-4}$ or better. This would allow to make use of the narrow-band
nature of harmonic lasing.

\section{Summary}

In this paper we have shown that harmonic lasing in X-ray FELs is a very attractive option.
It works well in the frame of the realistic 3D model of the FEL process, and can be realized in practice in a
relatively simple way, based on the achievements in production of high-brightness electron beams and high-quality
undulators. In this paper we have discussed only a few possible applications of harmonic lasing that can be summarized
as follows:

\begin{itemize}

\item extension of wavelength ranges of existing and planned X-ray FEL facilities beyond the baseline;

\item more flexible operation of facilities having several undulator beamlines;

\item reduction of the bandwidth at saturation (mild monochromatization) and an increase of brilliance;

\item simultaneous production of two colors for pump-probe experiments with an easy control of the intensity ratio;

\item fast switching between two or more colors for different user experiments;

\item a possibility of using more simple and robust undulator technology with larger periods and gaps at low-energy
X-ray FEL facilities.

\end{itemize}

We have also found out that harmonic lasing is, in principle, possible in soft x-ray beamlines of X-ray FEL facilities
without being aimed at. This lasing can be used as a mode of operation, or suppressed in a simple way.

Finally, let us express a hope that the list of possible applications will be significantly
extended in the future.

\section{Acknowledgments}
We would like to thank R. Brinkmann for his interest in this work, and D. Ratner for useful discussions.

\clearpage

\appendix

\section{Parametrization of the eigenvalue equation for harmonic lasing}

In Ref. \cite{ming} the eigenvalue equation for a high-gain FEL was derived that includes such important effects
as diffraction of radiation, betatron motion of particles and longitudinal velocity spread due to emittance,
energy spread in the electron beam, frequency detuning.
The eigenvalue equation is an integral equation which can be evaluated
numerically for any particular parameter set with a desirable accuracy. The generalization of
this eigenvalue equation to the case of harmonic lasing was done in \cite{kim-1}. Here we present the latter result
for the growth rate of $TEM_{nm}$ mode in a dimensionless form accepted in \cite{emit1}:

\begin{eqnarray}
\bar{\Phi}_{nm}(p)
& = &
-\frac{ h^2 A_{JJh}^2}{A_{JJ1}^2 (2 \I h B \hat{\Lambda} - p^2)}
\int \limits^{\infty}_{0} \D p' p'
\bar{\Phi}_{nm} (p') \nonumber \\
& \times &
\int \limits^{\infty }_{0} \D \zeta
\frac{\zeta}
{(1- \I h B \hat{k}_{\beta}^2 \zeta/2)^2}
\exp \left[ -\frac{
h^2 \hat{ \Lambda}^2_{\mathrm{T}} \zeta^2}{2} -
(\hat{\Lambda} + \I \hat{C}) \zeta \right] \nonumber \\
& \times &
\exp \left[ -\frac{p^2 + p'^2}{4(1- \I h B \hat{k}_{\beta}^2 \zeta/2)}
\right]
I_n \left[ \frac{ p p' \cos (\hat{k}_{\beta} \zeta)}
{ 2 (1- \I h B \hat{k}_{\beta}^2 \zeta/2)} \right] \ .
\label{eq:hank-trans}
\end{eqnarray}

\noindent where $h=1,3,5,...$ is harmonic number, $I_n$ is the modified Bessel function of the first
kind. The normalized growth rate $\hat{\Lambda} = \Lambda/\Gamma$ has to be found from numerical solution of
the integral equation.
The following notations are used here:  $\hat{r} = r/(\sigma
\sqrt{2})$, $B = 2 \sigma^{2}\Gamma \omega_1 /c$ is the diffraction
parameter, $\omega_1$ is the fundamental frequency,
$\sigma= \sqrt{\epsilon \beta}$ is the transverse rms size of the matched Gaussian beam,
emittance $\epsilon$ is simply given by $\epsilon = \epsilon_n/\gamma$
with $\epsilon_n$ being normalized rms emittance,
$\hat{k}_{\beta} = k_{\beta}/\Gamma$ is the
betatron motion parameter, $k_{\beta}=1/\beta$ is the betatron wavenumber, $\beta$ is the beta-function,
$\hat{\Lambda }^{2}_{\mathrm{T}} =
\sigma_{\gamma}^2/ (\bar{\rho} {\gamma})^2$ is the energy
spread parameter, $\hat{C} = \left[ k_{\mathrm{w}} - \omega_h/(2hc
\gamma_{z}^2) \right]/\Gamma$ is the detuning parameter, $\omega_h \simeq h \omega_1$, $\Gamma =
\left[ A_{\mathrm{JJ1}}^2 I \omega_1^{2}\theta
^{2}_{\mathrm{s}}\left(I_{\mathrm{A}}c^{2}\gamma ^{2}_{z}\gamma
\right)^{-1} \right]^{1/2}$ is the gain factor, $\bar{\rho} = c
\gamma_{z}^2 \Gamma/\omega_1$ is the efficiency parameter,
$\theta_{\mathrm{s}}=K/\gamma$, $K$ is
the rms undulator parameter, $\gamma$ is relativistic factor, $\gamma
^{-2}_{z} = \gamma ^{-2}+ \theta ^{2}_{\mathrm{s}}$, $k_{\mathrm{w}}$
is the undulator wavenumber, $I$ is the beam current, $I_{\mathrm{A}} =$ 17 kA is the Alfven
current, $A_{\mathrm{JJh}} = J_{(h-1)/2}(h K^2/2(1+K^2))
- J_{(h+1)/2}(h K^2/2(1+K^2))$.
Note that the scaling factors ($\Gamma$, $\bar{\rho}$) reflect the growth rate of the fundamental harmonic.
The efficiency parameter $\bar{\rho}$
is related to the corresponding parameter $\rho$ \cite{bon-rho} of the
one-dimensional model as follows: $\bar{\rho} = \rho B^{1/3}$.

One can observe that the
equation (\ref{eq:hank-trans})  can be rewritten such that it looks the same for all harmonics:

\begin{eqnarray}
\bar{\Phi}_{nm}(p)
& = &
-\frac{ 1}{ 2 \I \tilde{B} \tilde{\Lambda} - p^2}
\int \limits^{\infty}_{0} \D p' p'
\bar{\Phi}_{nm} (p') \nonumber \\
& \times &
\int \limits^{\infty }_{0} \D x
\frac{x}
{(1- \I \tilde{B} \tilde{k}_{\beta}^2 x/2)^2}
\exp \left[ -\frac{
\tilde{\Lambda}^2_{\mathrm{T}} x^2}{2} -
(\tilde{\Lambda} + \I \tilde{C}) x \right] \nonumber \\
& \times &
\exp \left[ -\frac{p^2 + p'^2}{4(1- \I \tilde{B} \tilde{k}_{\beta}^2 x/2)}
\right]
I_n \left[ \frac{ p p' \cos (\tilde{k}_{\beta} x)}
{ 2 (1- \I \tilde{B} \tilde{k}_{\beta}^2 x/2)} \right] \ ,
\label{eq:hank-trans-1}
\end{eqnarray}

\noindent with the following scaling factors:
$\tilde{\Gamma} =
\left[ A_{\mathrm{JJh}}^2 I \omega_h ^{2}\theta
^{2}_{\mathrm{s}}\left(I_{\mathrm{A}}c^{2}\gamma ^{2}_{z}\gamma
\right)^{-1} \right]^{1/2}$ and $\tilde{\rho} = c
\gamma_{z}^2 \tilde{\Gamma}/\omega_h$.
Note that the gain parameter can be rewritten as

\begin{equation}
\tilde{\Gamma} =
\left( \frac{A_{\mathrm{JJh}}^2 I\omega_h ^{2} K^2(1+K^2)}{I_{\mathrm{A}}c^{2}\gamma^5} \right)^{1/2}
\label{gain-param}
\end{equation}

The new scaled parameters are now written
as follows: $\tilde{\Lambda }^{2}_{\mathrm{T}} = \sigma_{\gamma}^2/(\tilde{\rho} {\gamma})^2$ is the energy spread parameter,
$\tilde{k}_{\beta} = k_{\beta}/\tilde{\Gamma}$ is the betatron motion parameter,
$\tilde{C} = \left[ k_{\mathrm{w}} - \omega_h/(2hc \gamma_{z}^2) \right]/\tilde{\Gamma}$ is the detuning parameter,
and

\begin{equation}
\tilde{B} = 2 \sigma^{2} \tilde{\Gamma} \omega_h /c
\label{diff-new}
\end{equation}

\noindent is the diffraction parameter.

In this paper we concentrate on the case when beta-function is optimized for the
highest FEL gain. Since diffraction parameter depends on beta-function,
it is more convenient to go over to the normalized parameters other then those introduced above.
Indeed, the diffraction parameter can be
rewritten as $\tilde{B} = 2 \tilde{\epsilon}/\tilde{k}_{\beta}$, where $\tilde{\epsilon}=2\pi\epsilon/\lambda_h$ and
$\lambda_h= 2\pi c/\omega_h$.
Then we can go from parameters $(\tilde{B},\tilde{k}_{\beta})$ to $(\tilde{\epsilon},\tilde{k}_{\beta})$,
and the Eq.~(\ref{eq:hank-trans-1}) becomes

\begin{eqnarray}
\bar{\Phi}_{nm}(p)
& = &
-\frac{ 1}{ 4 \I \tilde{\epsilon} \tilde{\Lambda}/\tilde{k}_{\beta} - p^2}
\int \limits^{\infty}_{0} \D p' p'
\bar{\Phi}_{nm} (p') \nonumber \\
& \times &
\int \limits^{\infty }_{0} \D x
\frac{x}
{(1- \I \tilde{\epsilon} \tilde{k}_{\beta} x)^2}
\exp \left[ -\frac{
\tilde{\Lambda}^2_{\mathrm{T}} x^2}{2} -
(\tilde{\Lambda} + \I \tilde{C}) x \right] \nonumber \\
& \times &
\exp \left[ -\frac{p^2 + p'^2}{4(1- \I \tilde{\epsilon} \tilde{k}_{\beta} x)}
\right]
I_n \left[ \frac{ p p' \cos (\tilde{k}_{\beta} x)}
{ 2 (1- \I \tilde{\epsilon} \tilde{k}_{\beta} x)} \right]  \ .
\label{eq:hank-trans-2}
\end{eqnarray}

Our goal is to find the reduced growth rate (the real part of the eigenvalue)
$Re \tilde{\Lambda} = Re \Lambda/\tilde{\Gamma}$ of the transverse mode $TEM_{00}$ when an FEL lases at
$h$-th harmonic. The field gain length of this mode is then simply $L_g = 1/ Re \Lambda$. In the case of a SASE FEL the
detuning parameter falls out of the parameters of the problem since the lasing always takes place at the
optimal detuning. Thus, when solving the eigenvalue equation, we should always find the eigenvalue at
the optimal detuning. Let us also assume at the first step that the energy spread parameter is negligibly small
(denoting the gain length for this case as $L_{g0}$), so that
its influence on FEL operation can be neglected. In this case the reduced growth rate $Re \tilde{\Lambda}$ depends only on
two dimensionless parameters: $\tilde{\epsilon}$ and $\tilde{k}_{\beta}$. If in addition one optimizes beta-function, then
the reduced growth rate is the function of the only parameter, scaled emittance: $Re \tilde{\Lambda} = f(\tilde{\epsilon})$.
Correspondingly, the field gain length can be written as follows:

\begin{equation}
L_{g0} = [ \tilde{\Gamma} f (\tilde{\epsilon})]^{-1}
\label{gain-l-1}
\end{equation}

Numerical solution of the eigenvalue equation (\ref{eq:hank-trans-2}) is time-consuming,
so we used an approximate solution \cite{emit1}
which agrees very well (to better than 1\% in the whole parameter space) with the solution of Eq.~(\ref{eq:hank-trans-2}).
In the most interesting parameter range, $1< \tilde{\epsilon} < 5$,
we have found \cite{des-form} that the function $f(\tilde{\epsilon})$ is well approximated as
$f(\tilde{\epsilon}) \propto \tilde{\epsilon}^{-5/6}$, so that the gain length in
the case of negligible energy spread and optimal beta-function is

\begin{equation}
L_{g0} \simeq a_1 \tilde{\Gamma}^{-1} \tilde{\epsilon}^{5/6} \ ,
\label{gain-l-2}
\end{equation}

\noindent where $a_1$ is the fitting coefficient. Now we would like to include the effects of the energy spread. For that
we present the growth rate as $L_g = L_{g0} (1+\delta)$, where $\delta$ depends on the energy spread. Again, for the optimal
beta-function, we found that the fit $\delta \propto \tilde{\Lambda }^{2}_{\mathrm{T}} \tilde{\epsilon}^{5/4}$
works very well in the wide range of values of the energy spread parameter. Thus, the field gain length for the
optimal beta-function can be written as follows:

\begin{equation}
L_g \simeq a_1 \tilde{\Gamma}^{-1} \tilde{\epsilon}^{5/6}
(1 + a_2 \tilde{\Lambda }^{2}_{\mathrm{T}} \tilde{\epsilon}^{5/4}) \ .
\label{gain-l-3}
\end{equation}

\noindent Optimizing fitting coefficients $a_1$ and $a_2$ in the range of parameters, specified in (\ref{emit-lam-lim}),
(\ref{delta-lim}), we obtain the Eqs.~(\ref{lg})-(\ref{delta}).
In a similar way we obtained the expression (\ref{beta}) for the optimal beta-function.
In particular, in the case of negligibly small
energy spread we used the following approximation: $(\tilde{k}_{\beta})_{opt} \propto \tilde{\epsilon}^{-3/2}$.

\section{Influence of quantum diffusion in an undulator on saturation length}

Energy spread growth due to the quantum fluctuations of the spontaneous
undulator radiation can be
an important effect \cite{qf-limit,qf-und} in X-ray FELs.
The rate of the energy diffusion is given by \cite{qf-und} (note that the peak value of the undulator parameter $K$ was
used in formulas of Ref.~\cite{qf-und}):

\begin{equation}
\frac {d \sigma_{\gamma}^2}{dz} =
\frac{14}{15} \lambar_{\mathrm{c}}r_{\mathrm{e}} \gamma ^4
\kappa _{\mathrm{w}}^3 K^2 F(K) \ ,
\label{eq:energy-diffusion-2}
\end{equation}

\noindent where $\lambar_{\mathrm{c}} = 3.86 \times 10^{-11}$ cm,
$r_{\mathrm{e}}=2.82 \times 10^{-13}$ cm,
$\kappa _{\mathrm{w}} = 2\pi/\lambda_{\mathrm{w}}$, and

\begin{equation}
F(K)  =  1.70 K + (1 + 1.88 K + 0.80 K^2)^{-1}
\label{fk}
\end{equation}

\noindent for planar undulator.
To estimate the FEL saturation length for the case of optimal beta-function,
we accept the following scheme \cite{des-form}. First, we neglect
energy diffusion and find a zeroth order approximation to the saturation length from
(\ref{lsat}), (\ref{lg})-(\ref{delta}). Then we calculate an induced energy spread
in the middle of the undulator
from (\ref{eq:energy-diffusion-2}), add it quadratically to the initial energy spread,
and find a new expression for $\delta$. Then, using (\ref{lsat}), (\ref{lg})-(\ref{delta}),
we find the first
approximation to the saturation length. Then we do the next iteration, etc. Finally,
the saturation length can be estimated as

\begin{equation}
L_{\mathrm{sat}} \simeq 10 \ L_{g0} \ \frac{1+\delta}{1-\delta_{q}} \ ,
\label{lsat-q}
\end{equation}

\noindent where

\begin{equation}
\delta_{q} = 5.5\times 10^4
\left(\frac{I_A}{I}\right)^{3/2} \frac{\lambar_{\mathrm{c}}r_{\mathrm{e}}\epsilon_n^2}
{\lambda_r^{11/4} \lambda_{\mathrm{w}}^{5/4}} \ \frac{(1+K^2)^{9/4}F(K)}{K A_{JJh}^3 h^{5/3}}
\label{delta-q}
\end{equation}

\noindent Note that in the latter formula the powers are somewhat simplified.
Comparing Eqs. (\ref{lsat}) and (\ref{lsat-q}), we can introduce an effective parameter

\begin{equation}
\delta_{\mathrm{eff}} = \frac{\delta+\delta_q}{1-\delta_q} \ ,
\label{delta-eff}
\end{equation}

\noindent which should be used instead of $\delta$ in (\ref{delta-lim}) to check
the applicability range and in (\ref{beta}) to estimate the optimal beta-function.

Although formula (\ref{lsat-q}) is rather crude estimate, it can be used for quick
orientation in the parameter space with {\it a posteriori} check using a numerical
simulation code.

\section{Generalization of Ming Xie formulas to the case of harmonic lasing}

In Refs.~\cite{ming-form-1,ming-form-2} the fitting formulas were presented that approximate
FEL {\bf power} gain length, $L_g$. Note that in
our parametrization in Section 2 (and throughout this paper) we use the same notation for the {\bf field} gain
length which is twice longer.
The power gain length of the fundamental harmonic was expressed in \cite{ming-form-1,ming-form-2} as follows:

\begin{equation}
\frac{L_{1d}}{L_g} = \frac{1}{1+\Lambda (\eta_d,\eta_{\epsilon},\eta_{\gamma})} \ ,
\label{ming-1}
\end{equation}

\noindent where $L_{1d}$ is the 1D gain length for the cold beam, and $\Lambda$ depends on the three dimensionless
parameters: $\eta_d$, $\eta_{\epsilon}$, and $\eta_{\gamma}$. This dependence can be found in
\cite{ming-form-1,ming-form-2}, it was obtained
by fitting the solution of the eigenvalue equation with the help of 19 fitting coefficients.

We can generalize these results for calculation of power gain length $L_g^{(h)}$ of harmonic lasing in a simple way.
Eq.~(\ref{ming-1}) can be generalized as

\begin{equation}
\frac{L_{1d}^{(h)}}{L_g^{(h)}} = \frac{1}{1+ \Lambda (\eta_d^{(h)},\eta_{\epsilon}^{(h)},\eta_{\gamma}^{(h)})} \ .
\label{ming-h}
\end{equation}

\noindent The 1D gain length of harmonics can be calculated as

\begin{displaymath}
L_{1d}^{(h)} = \left( \frac{A_{JJ1}^2}{h A_{JJh}^2} \right)^{1/3} L_{1d} \ ,
\end{displaymath}

\noindent and the function  $\Lambda$ now depends on the three generalized parameters:

\begin{displaymath}
\eta_d^{(h)} = \left( \frac{A_{JJ1}^2}{h A_{JJh}^2} \right)^{1/3}  \frac{\eta_d}{h} \ \ \ \ \ \ \ \
\eta_{\epsilon}^{(h)} = \left( \frac{A_{JJ1}^2}{h A_{JJh}^2} \right)^{1/3} h \eta_{\epsilon} \ \ \ \ \ \ \ \
\eta_{\gamma}^{(h)} = \left( \frac{A_{JJ1}^2}{h A_{JJh}^2} \right)^{1/3} h \eta_{\gamma}
\end{displaymath}

\section{Phase shifters method: 1D model}

It was suggested in \cite{mcneil} that the fundamental mode can be disrupted by introducing consecutive phase shifts $2\pi/3$
while the third harmonic is amplified without interruptions up to saturation. However, the simulations in \cite{mcneil}
were done for the case of a monochromatic seed. We would like to check if this method also works in the case of a SASE FEL,
using the same 1D model as in \cite{mcneil}, and similar normalization procedure.
For example, the reduced longitudinal coordinate $\hat{z}$ in our notations \cite{book}
corresponds to $\bar{z}$ in \cite{mcneil}.

\begin{figure*}[tb]

\includegraphics[width=.8\textwidth]{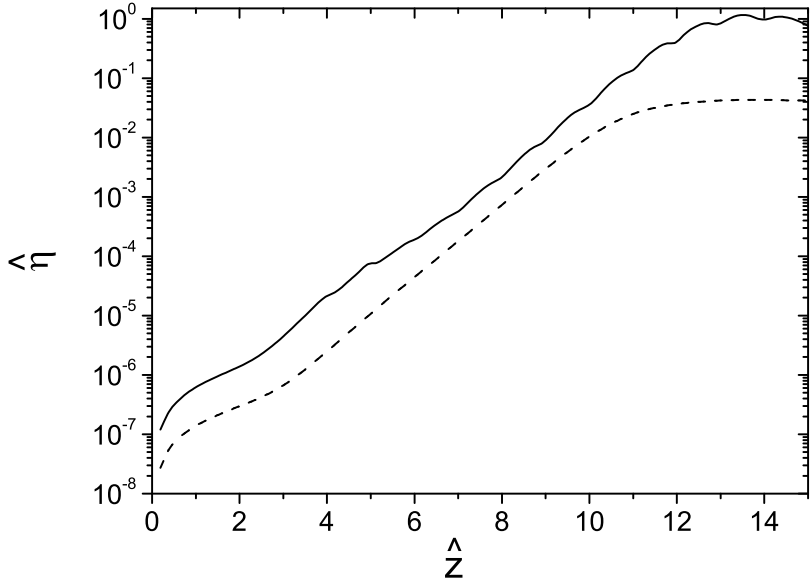}

\caption{\small Normalized power of the fundamental harmonic (solid) and of the third harmonic (dash) versus normalized
undulator length for a SASE FEL.
Phase shifts are equal to $2\pi/3$ at the positions $\hat{z} = 4,5,6,...,14$, as suggested in \cite{mcneil}.
The undulator parameter is large, $K \gg 1$. Definitions of the normalized parameters can be found in \cite{book}
}

\label{mcneil-fig}
\end{figure*}

We define phase shift in the same way as it was done in \cite{mcneil} to make the results compatible.
For example, the shift $2\pi/3$ corresponds to the
advance of a modulated electron beam w.r.t. electromagnetic field by $\lambda_1/3$. In Fig.~\ref{mcneil-fig} we present
the simulation of SASE FEL with the set of phase shifters considered in \cite{mcneil}:
phase shifts are equal to $2\pi/3$ at the positions $\hat{z} = 4,5,6,...$ . One can see that
this set does not provide a sufficient disruption of the fundamental mode so that it reaches saturation not allowing the
third harmonic to achieve high intensity level (although it is somewhat larger than that in the case without phase shifters).
Note that starting with phase shifts earlier, at $\hat{z}=1$, or using them more often does not bring a significant
improvement of the situation. Better results are achieved if one uses $4\pi/3$ shifts, but this is also not sufficient for
a sure suppression of the fundamental harmonic and obtaining an ultimate performance of the third harmonic. The main difference
of a SASE FEL with a seeded FEL amplifier is that in the former case the amplified frequency band is defined
self-consistently, i.e. the mean frequency is shifted depending on magnitude and positions of phase shifts.

\begin{figure*}[tb]

\includegraphics[width=.8\textwidth]{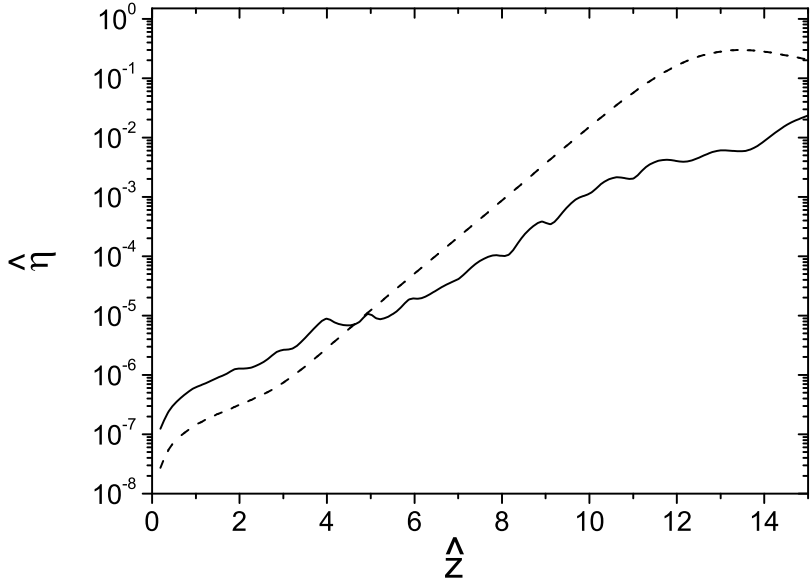}

\caption{\small Normalized power of the fundamental harmonic (solid) and of the third harmonic (dash) versus normalized undulator length.
Phase shifts are equal to $4\pi/3$ at the positions $\hat{z} = 1,2,3,8,9$ and
$2\pi/3$ at the positions $\hat{z} = 4,5,6,7,10,11$. The undulator parameter is large, $K \gg 1$.
}

\label{shifts}
\end{figure*}

If, however, we apply a modified method described in Section 4.1, namely
a picewise use of phase shifts $2\pi/3$ and $4\pi/3$,
we can achieve a desirable situation as one can see from Fig.~\ref{shifts}. In this case the third harmonic saturates while
the fundamental mode stays well below saturation.

\section{Harmonics versus the retuned fundamental mode:
comparison with 1D model}

Here we consider harmonic lasing and lasing {\bf at the same wavelength} at the fundamental mode
with the reduced undulator parameter $K$. In this Section we neglect energy
spread effects ($\delta=0$), and concentrate on comparison of 3D and 1D models.

A comparison between gain length of the third harmonic and that of the fundamental harmonic (with reduced
undulator parameter $K_{re}$ such that the wavelength is the same in both cases) was done in \cite{mcneil}
in the framework of 1D model.
It was shown that in the case of cold electron beam the FEL gain length is always shorter for third harmonic lasing.
The ratio
of gain length of the retuned fundamental mode to the gain length of the h-th harmonic is given by the formula:

\begin{equation}
\frac{L_{g}^{(1K)}}{L_{g}^{(h)}} = \left[ \frac{h K^2 A_{JJh}^2(K)}{K_{re}^2 A_{JJ1}^2(K_{re})} \right]^{1/3}
\ \ \ \ \ \ \ \ \ \ \ \ \ \ \ \ \ \ \ \ \ \ \ \ \  {\mathrm in \ 1D \ model}
\label{ratio-1h-1d}
\end{equation}

\noindent The superscript $(1K)$ indicates that the retunig of the undulator parameter was used to reduce wavelength
of the first harmonic.
The retuned undulator parameter $K_{re}$ can be found from the equation:

\begin{equation}
\frac{1+K^2}{1+K_{re}^2} = h
\label{ret-k}
\end{equation}

Now we can present the corresponding ratio of gain lengths for the case of a full 3D model of FEL process including diffraction
of radiation, finite transverse beam size, betatron motion etc. As it was done above, we assume that
beta-function is optimized in each case, and the energy spread effects can be neglected. Since the wavelength and
the beam energy in the considered case are the same for a harmonic and for the fundamental mode, the emittance parameter
$\tilde{\epsilon}$ is also the same. Thus, according to (\ref{gain-l-1}), the ratio of gain lengths is simply given
by the inverse ratio of gain parameters $\tilde{\Gamma}$ (see (\ref{gain-param})), i.e. it can be written
(with the help of (\ref{ret-k})) as follows:

\begin{equation}
\frac{L_{g}^{(1K)}}{L_{g}^{(h)}} =   \frac{h^{1/2} K A_{JJh}(K)}{K_{re} A_{JJ1}(K_{re})}
\ \ \ \ \ \ \ \ \ \ \ \ \ \ \ \  {\mathrm in \ 3D \ model}
\label{ratio-1h-3d}
\end{equation}

\begin{figure*}[t]

\includegraphics[width=.8\textwidth]{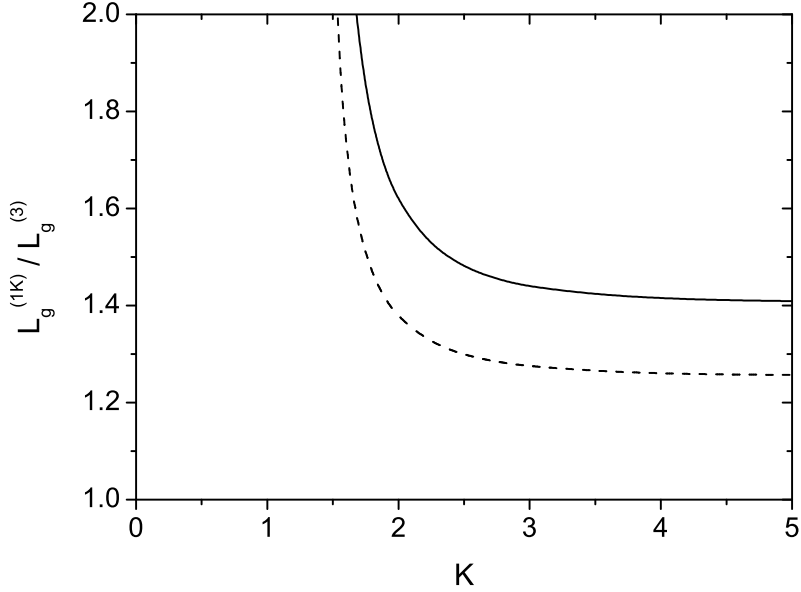}

\caption{\small Ratio of gain lengths for lasing at the fundamental wavelength and at the third harmonic. Solid curve is
calculated with the help of Eq.~(\ref{ratio-1h-3d}) in the frame of 3D model. Dashed curve is
calculated with the help of Eq.~(\ref{ratio-1h-1d}) in the frame of 1D model. Adjustment of the fundamental wavelength was done by
retuning of the undulator parameter K according to (\ref{ret-k}).
}

\label{1d-3d}
\end{figure*}

One can observe that Eqs.~(\ref{ratio-1h-1d}) and (\ref{ratio-1h-3d}) can be directly compared:

\begin{equation}
\left[ \frac{L_{g}^{(1K)}}{L_{g}^{(h)}} \right]_{3D,  \beta_{opt}} =
\left[ \frac{L_{g}^{(1K)}}{L_{g}^{(h)}} \right]_{1D}^{3/2}
\label{1d-3d-32}
\end{equation}

We indicated explicitly that in 3D case the beta-function was optimized for a harmonic and for the retuned fundamental.
In contrast, in 1D case the emittance and the beta-function are not the parameters of the problem,
and the current density is kept the same. It was shown in \cite{mcneil} that in the frame of 1D theory with cold electron beam
the gain length of the retuned fundamental mode is
always larger than the gain length of the third harmonic. It follows from (\ref{1d-3d-32}) that this also holds in 3D case
with negligible energy spread and optimal beta-function, moreover the ratio is even larger (it is raised to the power 3/2)
than in 1D case. This means that inclusion of 3D effects actually improves the situation
and makes harmonic lasing even more attractive option than 1D theory suggests, see Fig.~\ref{1d-3d}.

\end{document}